\documentclass[letterpaper, 10 pt, conference]{ieeeconf}  
\IEEEoverridecommandlockouts                              

\usepackage{graphics} 
\usepackage{epsfig} 
\usepackage{mathptmx} 
\usepackage{times} 
\usepackage{amsmath} 
\usepackage{amssymb}  
\usepackage[ruled]{algorithm2e}
\usepackage{subfigure}
\usepackage{tabu}
\usepackage{booktabs}
\usepackage{multirow}
\usepackage{footnote}
\usepackage{threeparttable}
\usepackage{booktabs}
\usepackage[table]{xcolor}
\usepackage{geometry}
\usepackage{tabularx}
\usepackage{tabularx}
\usepackage{array}

\author{Boyi Liu$^{{1},{3}{*}}$, Bingjie Yan$^{{2}{*}}$, Yize Zhou$^{2}$, Yifan Yang$^{2}$ and Yixian Zhang$^{2}$
	\thanks{$^{1}$Boyi liu is with Cloud Computing Lab of Shenzhen Institutes of Advanced Technology, Chinese Academy of Sciences. {\tt\small by.liu@ieee.org};
	{\tt\footnotesize lj.wang1@siat.ac.cn}}
	\thanks{$^{2}$Bingjie Yan is with Shool of Computer Science and Cyberspace Security, Hainan University. {\tt\footnotesize beiyuouo@foxmail.com}}
	\thanks{$^{3}$Boyi Liu are also with the University of Macau.}
}

\title{\LARGE \bf
	Experiments of Federated Learning for COVID-19 Chest X-ray Images
}
\begin{document}
	\maketitle
	\begin{abstract}
	AI plays an important role in COVID-19 identification. Computer vision and deep learning techniques can assist in determining COVID-19 infection with Chest X-ray Images. However, for the protection and respect of the privacy of patients, the hospital's specific medical-related data did not allow leakage and sharing without permission. Collecting such training data was a major challenge. To a certain extent, this has caused a lack of sufficient data samples when performing deep learning approaches to detect COVID-19. Federated Learning is an available way to address this issue. It can effectively address the issue of data silos and get a shared model without obtaining local data. In the work, we propose the use of federated learning for COVID-19 data training and deploy experiments to verify the effectiveness. And we also compare performances of four popular models (MobileNet, ResNet18, MoblieNet, and COVID-Net) with the federated learning framework and without the framework. This work aims to inspire more researches on federated learning about COVID-19.
	\end{abstract}

	\section{INTRODUCTION}
	The COVID-19 pandemic has caused continuous damage to the health and normal production of people all over the world. Therefore, researches on detecting and diagnosing COVID-19 patients are very meaningful\cite{yan2020improved}\cite{zhang2020covid}. The clinical manifestations of COVID-19 infected pneumonia are mainly fever, chills, dry cough, and systemic pain. A few patients have abdominal symptoms. It is worth noting that there are asymptomatic patients in the population. So it is necessary to test more people as soon as possible. A key step in judging and treating COVID-19 is the effective screening of infected patients. One of the key screening methods is the use of chest X-rays for radiology. Computer vision and machine learning technology play an important role in this approach. At present, artificial intelligence, especially deep learning, has become an important technology for computer-aided medical applications and has achieved remarkable results in medical imaging. Deep learning has made a huge contribution to the classification of chest X-ray radiology in the medical field, and it has become an effective tool for doctors to judge and analyze the condition. To obtain an accurate and robust depth model, the core element is large and widely diverse training data. However, out of the protection and respect of the privacy of patients, the hospital's specific medical-related data did not allow leakage and public research.
	\begin{figure}[!htbp]
		\centering
		\includegraphics[width=0.5\textwidth]{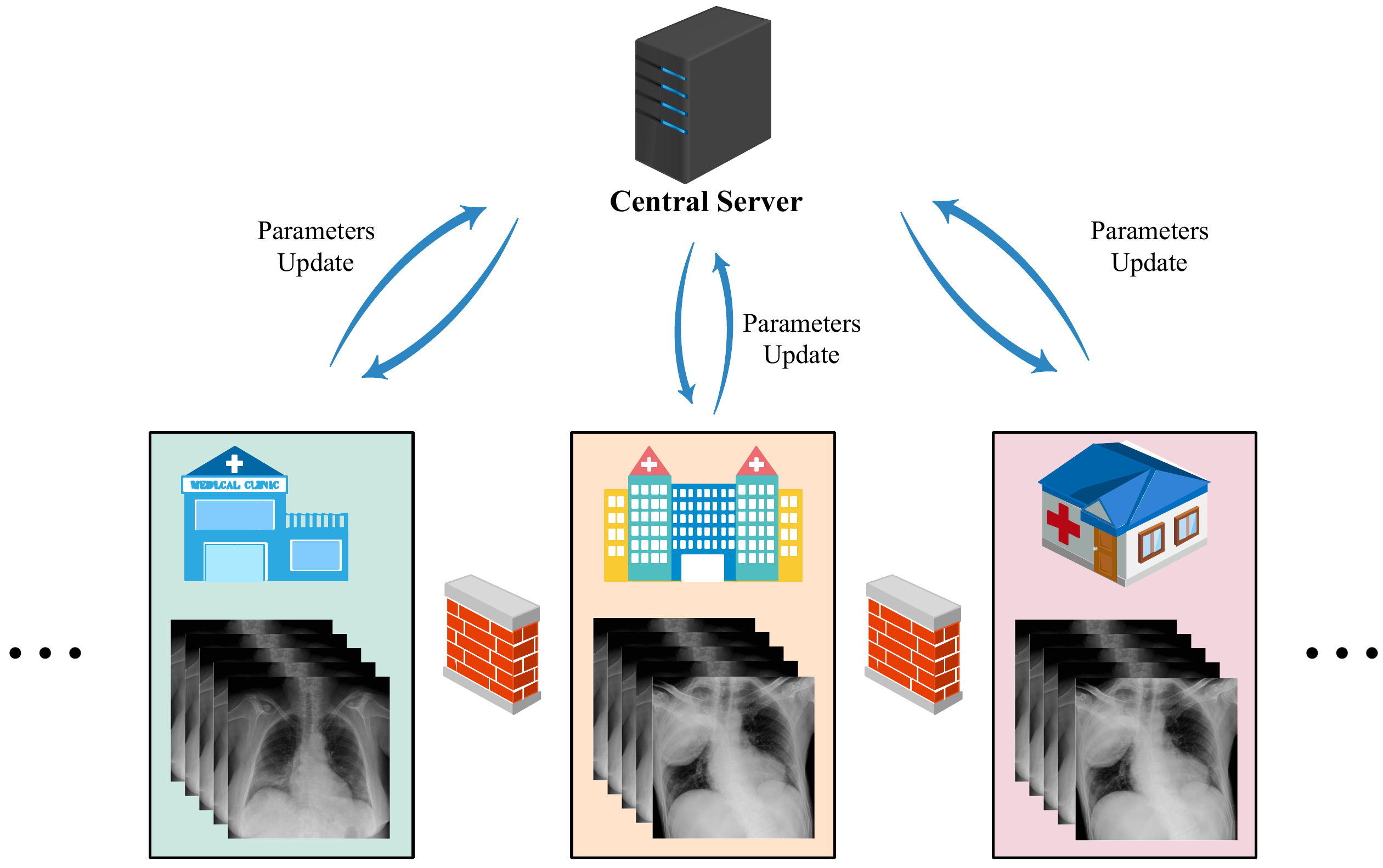}
		\caption{Federated Learning Framework for COVID-19 CXR images}
		\label{fig}
	\end{figure}
	Collecting such training data was a major challenge. To a certain extent, this has caused a lack of sufficient data samples when performing deep learning approaches to detect COVID-19. Federated Learning is an available way to address this issue. It can effectively address the issue of data silos and get a shared model without obtaining local data. In the paper, we firstly propose the use of federated learning for COVID-19 data training and deploy experiments to verify the effectiveness.

	Federated learning is a framework of learning across multiple institutions without sharing patient data. It has the potential to fundamentally solve the problems of data privacy and data silos. Applications of federated learning in medical big data are promising researches. Federated learning is capable of utilizing the non-shared data from different hospitals, enlarging the sample size of the model training, and improving the accuracy of the model. The core of federated learning is to use data sets distributed on multiple devices to jointly build a shared model and does not require local raw data sharing. This precisely protects patient data. In the case that COVID-19 medical imaging data is still distributed in various countries and hospitals, federated learning experiments for medical images of COVID-19 that conducted in this work are necessary.
	
	In this work, we conducted four individual experiments to present the performances in federated learning of four different networks for COVID-19 CXR images: CovidNet\cite{wang2020covid}, ResNeXt\cite{sharma2018spatial}, MobileNet-v2\cite{ sandler2018mobilenetv2}, and ResNet18\cite{ayyachamy2019medical}. Further, we analyzed the results and proposed possible future improvements to inspire more research in federated learning for COVID-19.
	\section{Related Work}
	\subsection{COVID-Net for COVID-19 Indentification}
	With the continuous development of the new coronavirus epidemic, more and more researchers are committed to join the ranks of fighting the epidemic through AI-related technologies. Researchers use AI to make it play a role in the epidemic. A series of recent studies on COVID-19 medical imaging analysis and diagnosis have sprung up. These studies have completed the diagnosis of COVID-19 based on medical imaging technology. We use the current CT scanning technology to complete the image generation of the special radiological features and image modes of COVID-19. After that, the researchers used machine learning methods to classify and recognize the images generated during the CT scan diagnosis process. This method greatly reduces the workload of medical staff, and at the same time plays a role in assisting doctors in diagnosing the pathological characteristics of patients. At present, many studies have targeted disease as a binary classification problem, that is, "health" and "positive new coronavirus". Next, we introduce the application of COVID-Net in COVID-19 image classification and recognition and complete diagnosis of pathological features.
	
	COVID-Net specifically proposes a neural network that uses PEPX compression network structure to identify COVID-19 pneumonia CXR images. At the same time it retains the performance of the network to a great extent and is highly sensitive to the pneumonia characteristics of COVID-19. Based on the advantages of CXR imaging for rapid triage of COVID-19 screening, availability, ubiquity, and portability, they make predictions through the COVID-Net interpretability method. This allows us to not only gain a deeper understanding of the key factors associated with COVID cases. This can help clinicians perform better screening. We can also review COVID-Net, a method based on CXR images to verify that it is making decisions.
	
	\subsection{Federated Learning}
	Federated learning is a recent emerging research that has been extensively studied in the fields of financial security, artificial intelligence, and robotics\cite{liu2019lifelong}\cite{liu2019federated}\cite{chen2020design}.
	The training data will be distributed on each mobile device, not all of them will be sent to the central server, and only the updated data on each device will be aggregated to the central server. After joint optimization, the central server returns to the global state of each device, and continues to accept the updated data calculated by each client in the new global state. This method is Federated Learning\cite{brendan2016communication}
	. Federated Learning or Federated Machine Learning\cite{konevcny2016federated} can solve the problem of unprotected large-scale private data and complete updating learning of devices without exchanging large amounts of data.
	
	This decentralized training model approach provides privacy, security, regulation, and economic benefits\cite{zhao2018federated}. Federated Learning presents new statistical and system challenges when training machine models on distributed device networks\cite{smith2017federated}. Federated Learning, which relies on scattered data, brings many aspects of research: Fei Chen et al. identified the combination of Federated Learning and Meta-learning as a major advance in Federated Learning\cite{chen2018federated}. Konstantin Sozinov et al have made some progress in applying Federated Learning to human activity identification\cite{sozinov2018human}.
	\section{Federated learning System for COVID-19 CXR Images}
	In this section, we provide a comprehensive overview of federated learning. Furthermore, the definition, architecture, training process and parameters update method of the federated learning system [5] for COVID-19 CXR images are considered.
	
	\subsection{Basic Definition}
	In the work, we define N COVID-19 CXR images owners as ${F_1,F_2,……F_N}$. We assume that they are from different hospitals. Patient medical data is not allowed to be shared, including CXR images. Under this constraint, all of them want to train their own model by merging their respective data ${D_1,D_2……D_N}$. A conventional method exists to put all the data together and use $D=D_1 \bigcup D_2 ……D_N$ to train to get a model $M_SUM$. Federated learning is a systematic learning process. In this process, the data owners jointly train the model $M_FED$. During this process, any data owner $F_i$ will not disclose their own data $D_i$ to others. In addition to this, the accuracy of $M_FED$ represented as $V_FED$ should be very close to the $V_SUM$ performance of $M_SUM$. In the form of expression, let $\epsilon$ be a non-negative real number; If $\arrowvert V_FED-V_SUM \arrowvert <\epsilon$, we can think that the federated learning algorithm has a $\epsilon$ error accuracy.
	
	\subsection{Framework of the Federated Learning System}
	In this part, we will introduce the basic framework of federated learning, the training structure, and the way to update the parameters. Federated learning is a distributed learning method. The server is used to maintain the overall main model and distribute it to various user terminals. For privacy issues, users train learning models on local terminals. The server will set the score S, and extract the user terminal according to the proportion to update the central model of the server. Then upload the user-improved model parameters to the server to update the server model parameters. Subsequently, it is distributed to user terminals to improve the user terminal model. In this way, we continue to improve the central model of the server and the local model of the user terminal. This approach is capable of ensuring the accuracy and privacy of the user terminal,  utilize the user terminal's computing power and a large amount of user data to learn, and maintain an excellent central model.
	
	In the FL training system, the owner of the data acts as a participant in the FL process. And they jointly train the machine learning (ML) model of the aggregation server center. In this control architecture, a basic premise assumption is that the data owners are honest and the data they provide is true. This requires data users to use their real private data for training and submit the trained local model to the FL server.
	
	Generally, the FL training process includes the following three training steps. We first define that the local model refers to the model trained on each participating device, and the global model refers to the model after the FL server has been aggregated.
	
	\begin{itemize}
		\item Step 1: Implements task initialization. The server determines the training task, which is to determine the target application and corresponding data requirements. At the same time, the server specifies the global model and establishes parameters during the training process, such as the learning rate. Afterwards, the server allocates the initialized global model $W_{G}0$ and training tasks to the participating clients to complete the task allocation.
		\item Step 2: Implements the training and update of the local model. The training is carried out on the basis of the global model $W_{t}G$, where t represents the current iteration index, and each participating user uses local data and equipment to update the local model parameters $W_{t_i}$. The final goal of participant i in iteration t is to find the optimal parameter $W_{t_i}$ that minimizes the loss function $L(W_{t_i})$.
		\item Step 3: Realizing the aggregation and update of the global model. The server aggregates the local models of the participating users and sends the updated global model parameters $W_(t+1)$ G to the users who hold the data.
	\end{itemize}
	\section{Experiments}
	In this session, we will explain our experiments on the recognition of COVID-19 pneumonia CXR images using various models and federated learning frameworks.
	\subsection{Dataset}
	The dataset used to train and evaluate model is COVIDx dataset, which is one of the open access dataset with the largest number of COVID-19 pneumonia CXR images. It contains covid chestxray dataset, COVID-19 Chest X-ray Dataset, Actualmed COVID-19 Chest X-ray Dataset, covid-19 radiography dataset which is public on kaggle and RSNA Pneumonia Detection Challenge’s dataset. There are 15,282 images in this dataset, including 13,703 images for training and 1,579 images for testing. There are three kinds of labels in the dataset. They are Normal (which is asymptomatic), pneumonia (which is non-COVID19 pneumonia) and COVID19 (which is pneumonia caused by novel coronavirus). The various data distributions are shown in Figure 1.
	\begin{figure}[!htbp]
		\centering
		\includegraphics[width=0.5\textwidth]{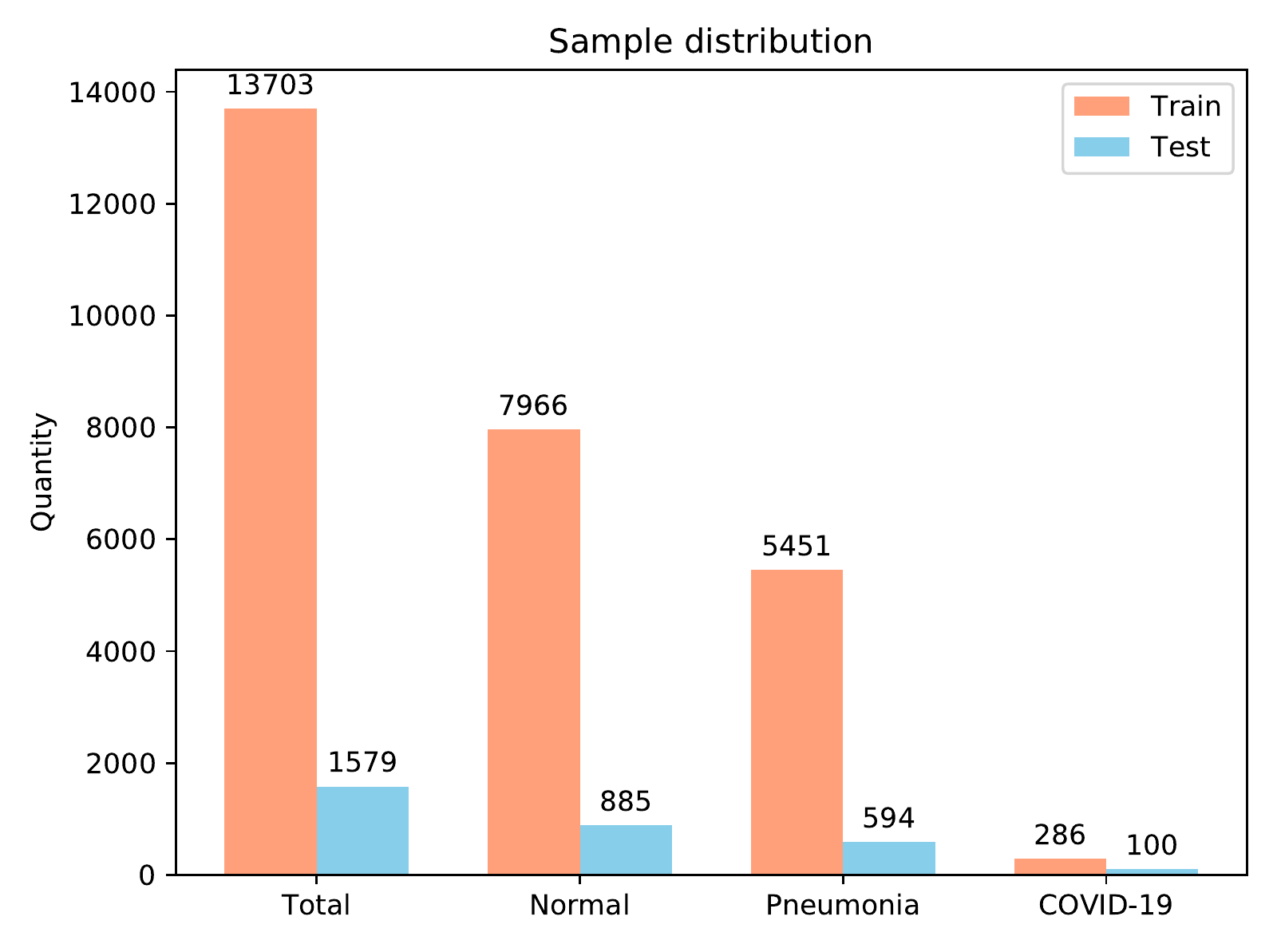}
		\caption{Data distributions in COVIDx dataset}
		\label{fig}
	\end{figure}
	\begin{figure*}
		\centering
		\includegraphics[width=0.5\textwidth]{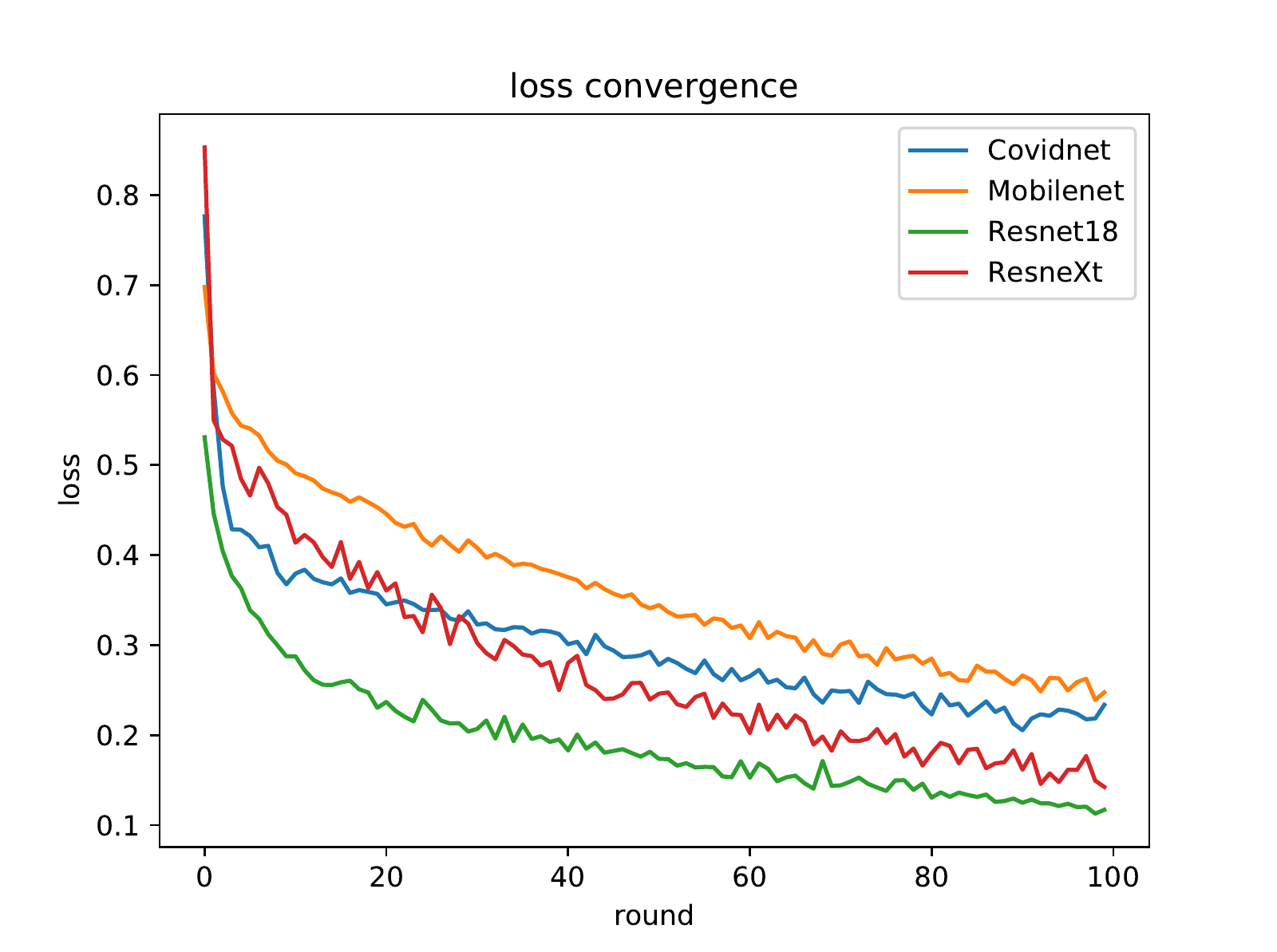}
		\caption{Loss convergence with training rounds}
		\label{fig}
	\end{figure*}
	\begin{figure*}
		\centering
		\subfigure[CovidNet FL]{\includegraphics[width=3.2in]{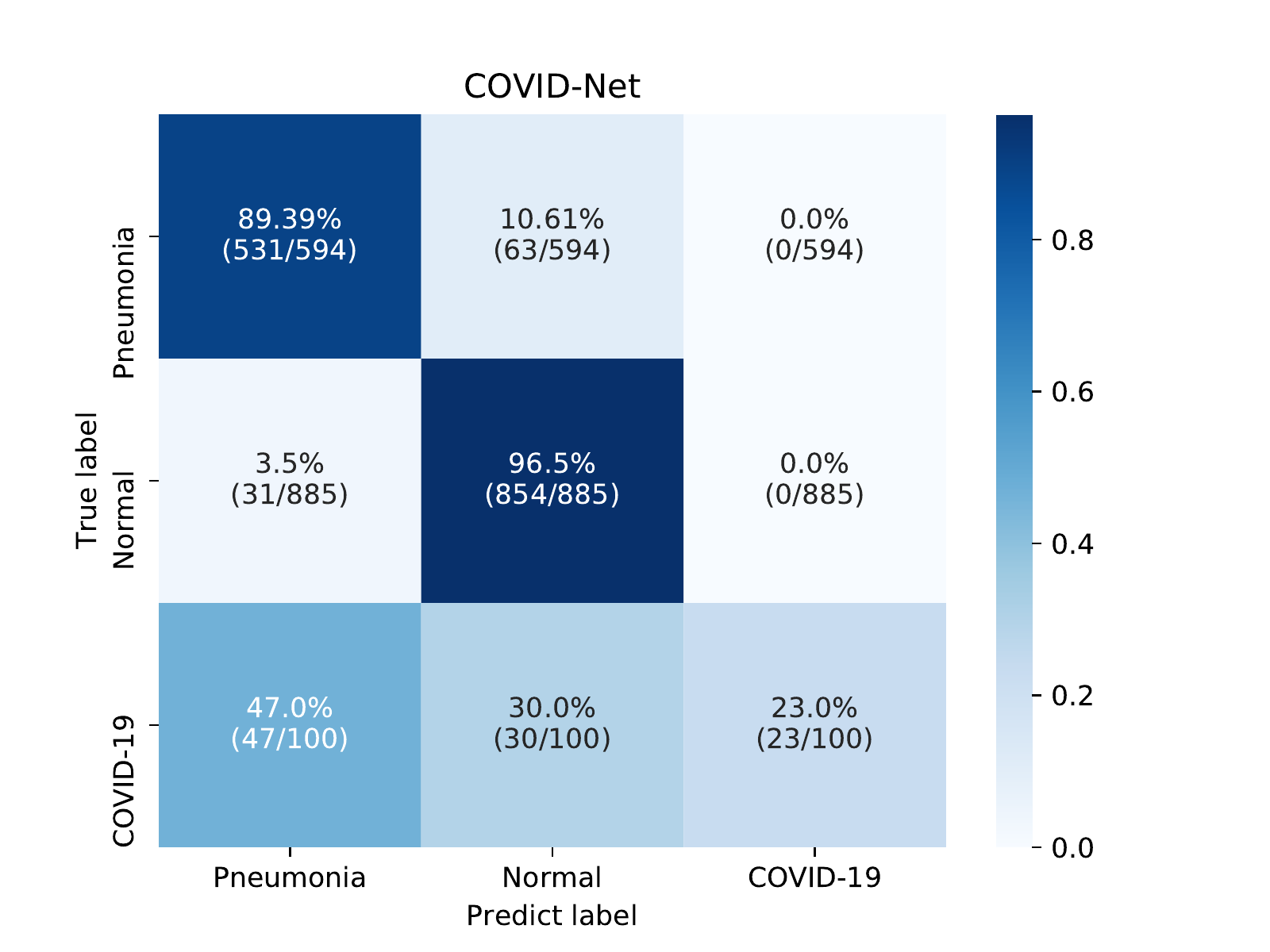}}
		\subfigure[MobileNet2 FL]{\includegraphics[width=3.2in]{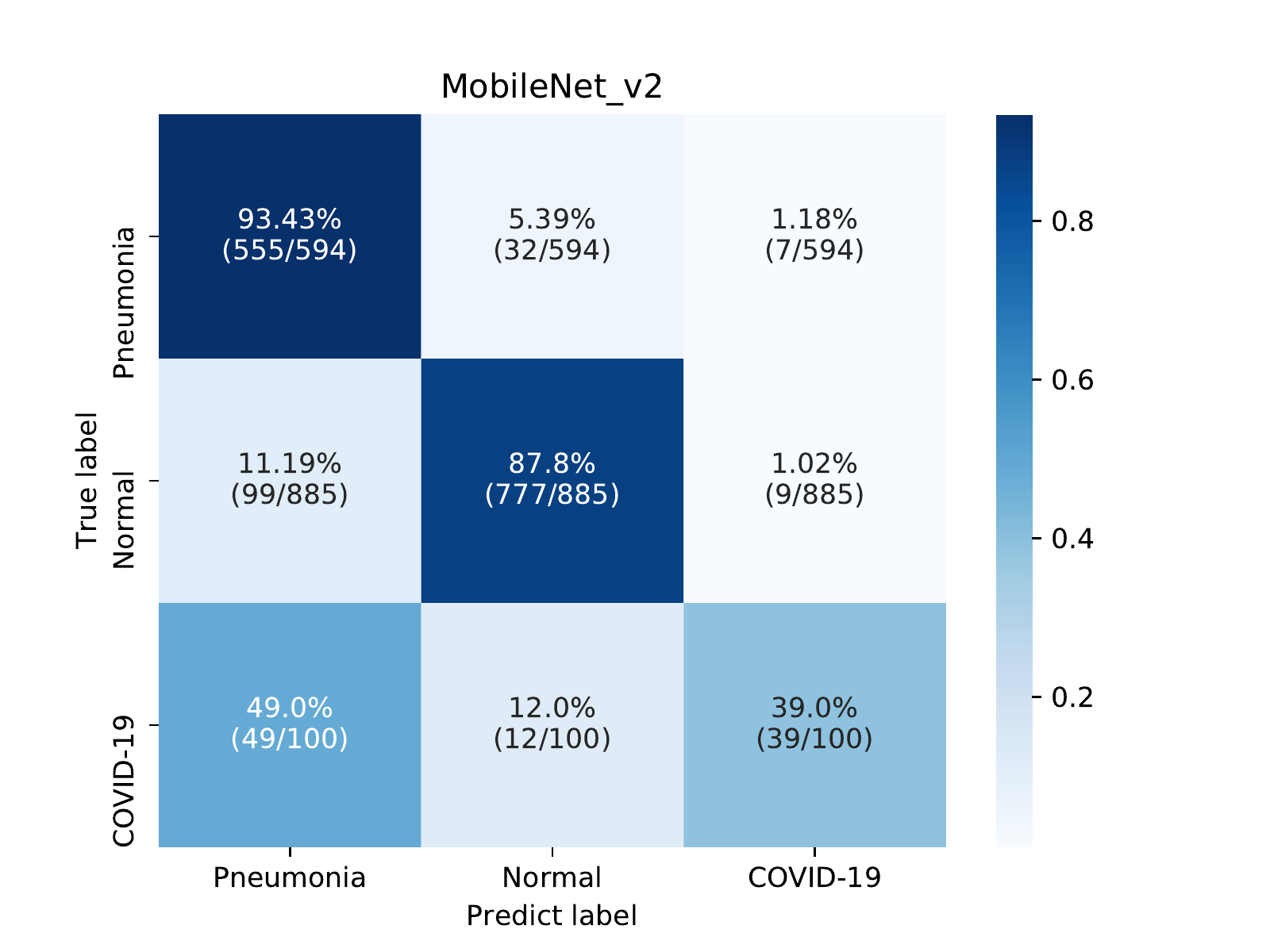}}
		\subfigure[ResNet18 FL]{\includegraphics[width=3.2in]{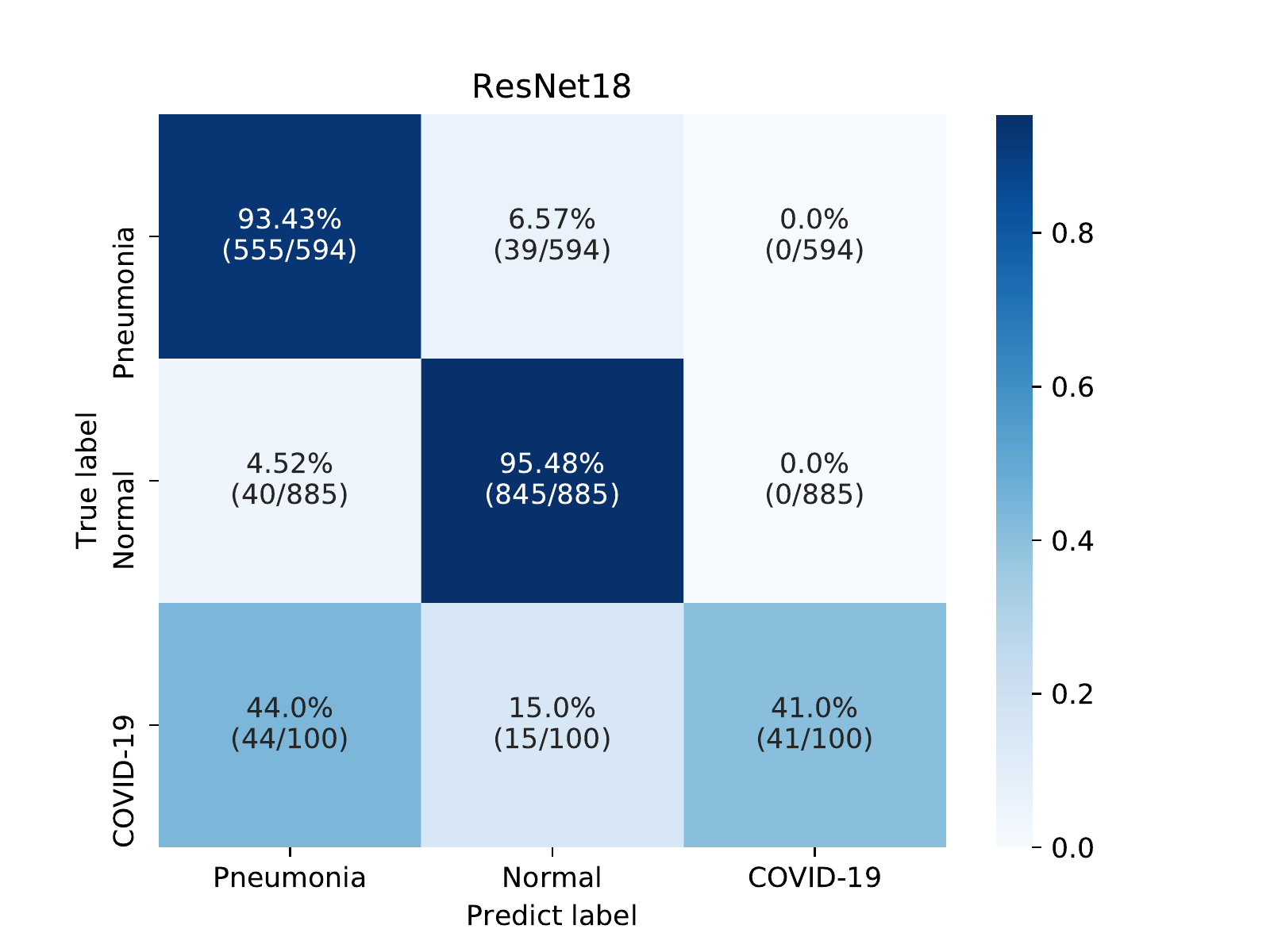}}
		\subfigure[ResNeXt FL]{\includegraphics[width=3.2in]{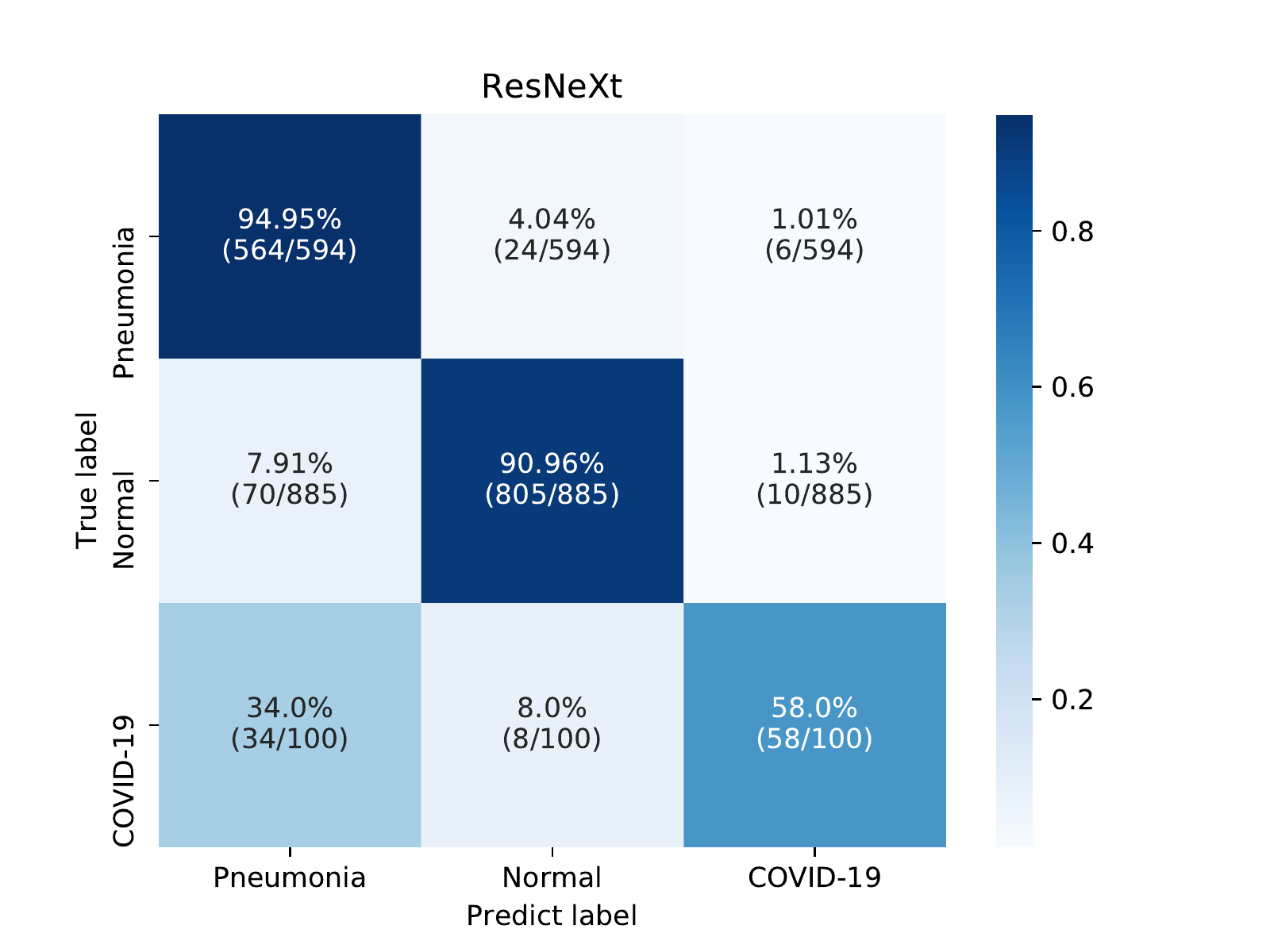}}
		\caption{Four models perplexity of each label}
	\end{figure*}
		\begin{figure*}
		\centering
		\subfigure[CovidNet FL]{\includegraphics[width=3.2in]{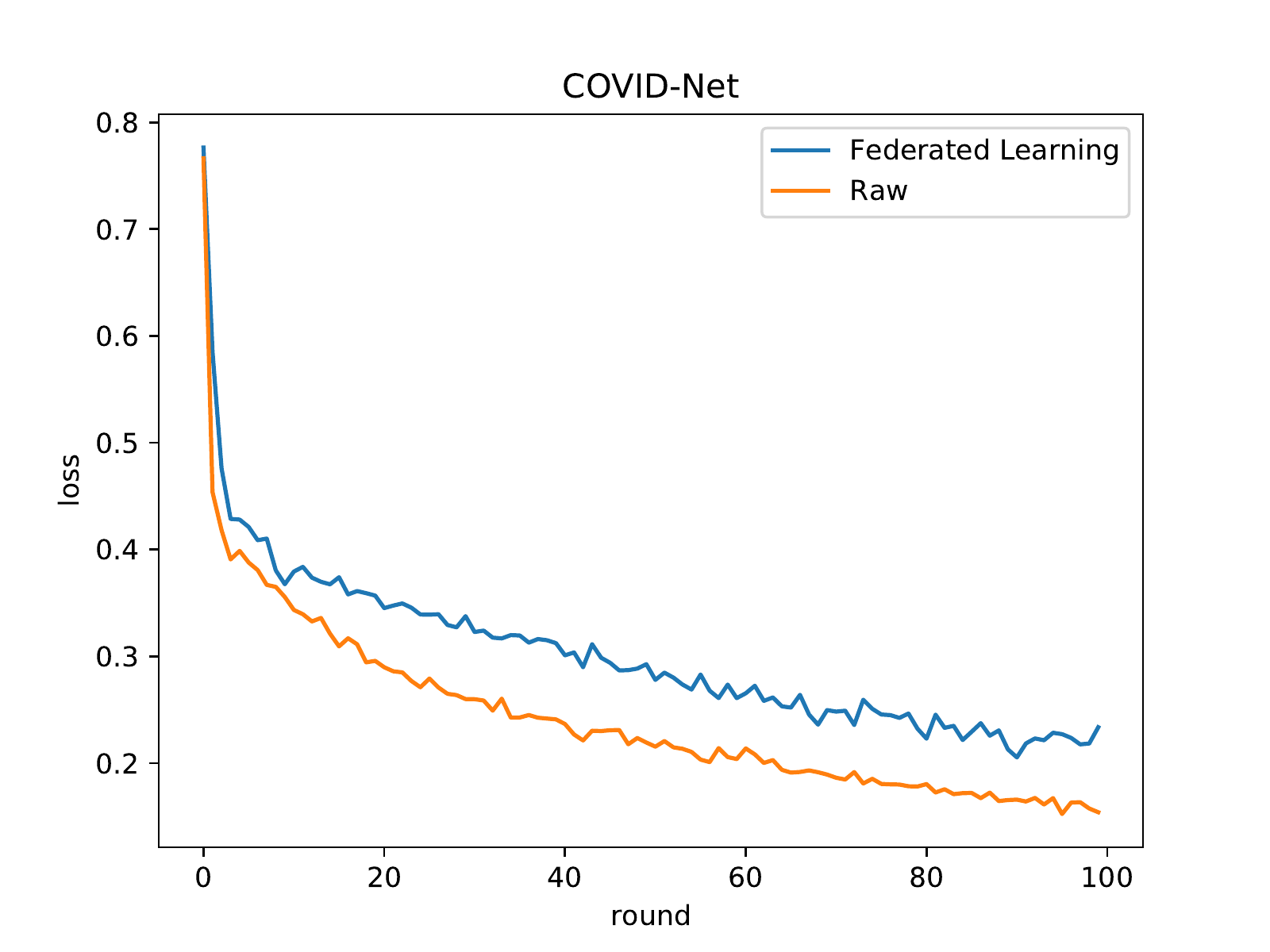}}
		\subfigure[MobileNet-v2 FL]{\includegraphics[width=3.2in]{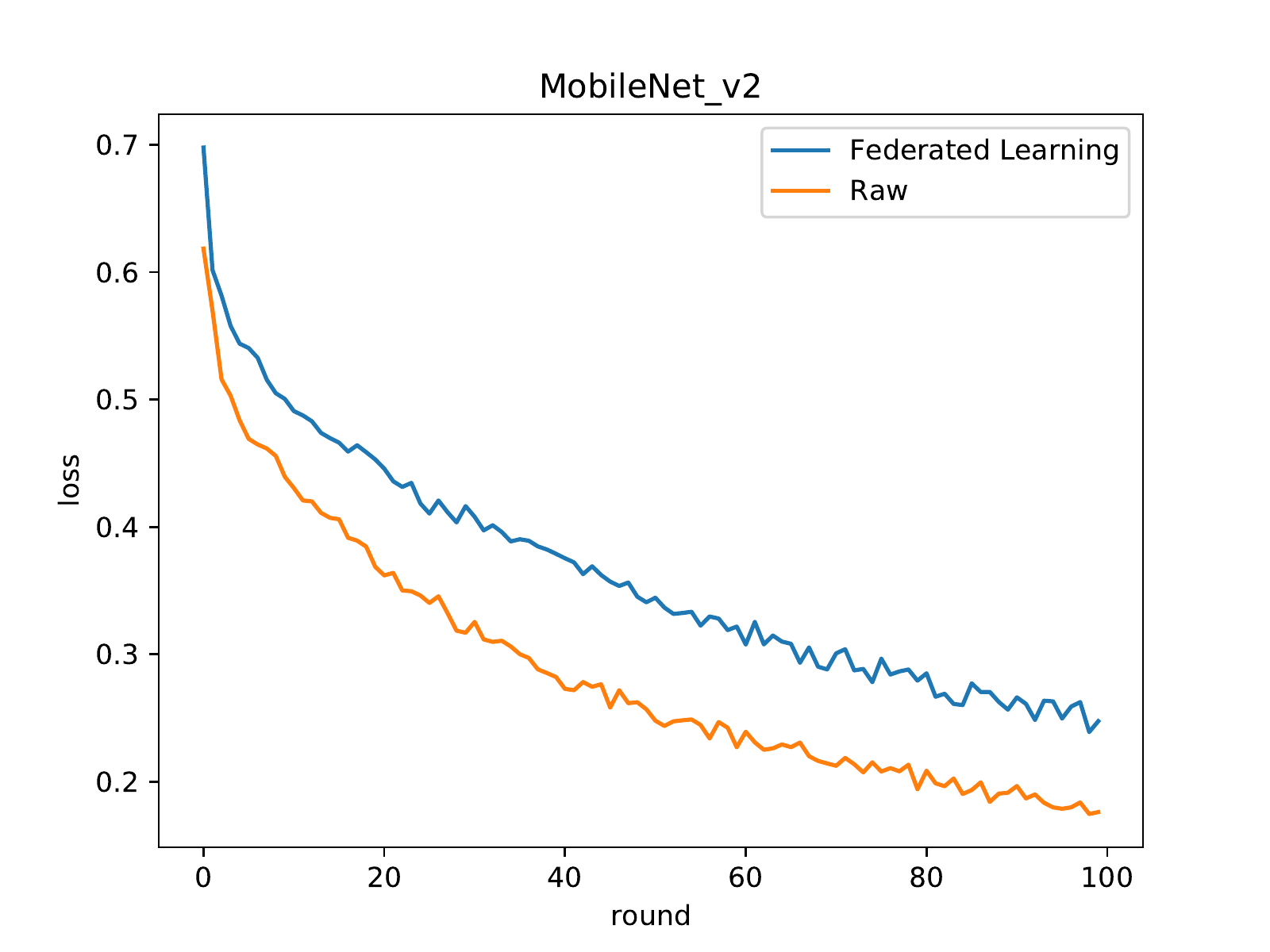}}
		\subfigure[ResNet18 FL]{\includegraphics[width=3.2in]{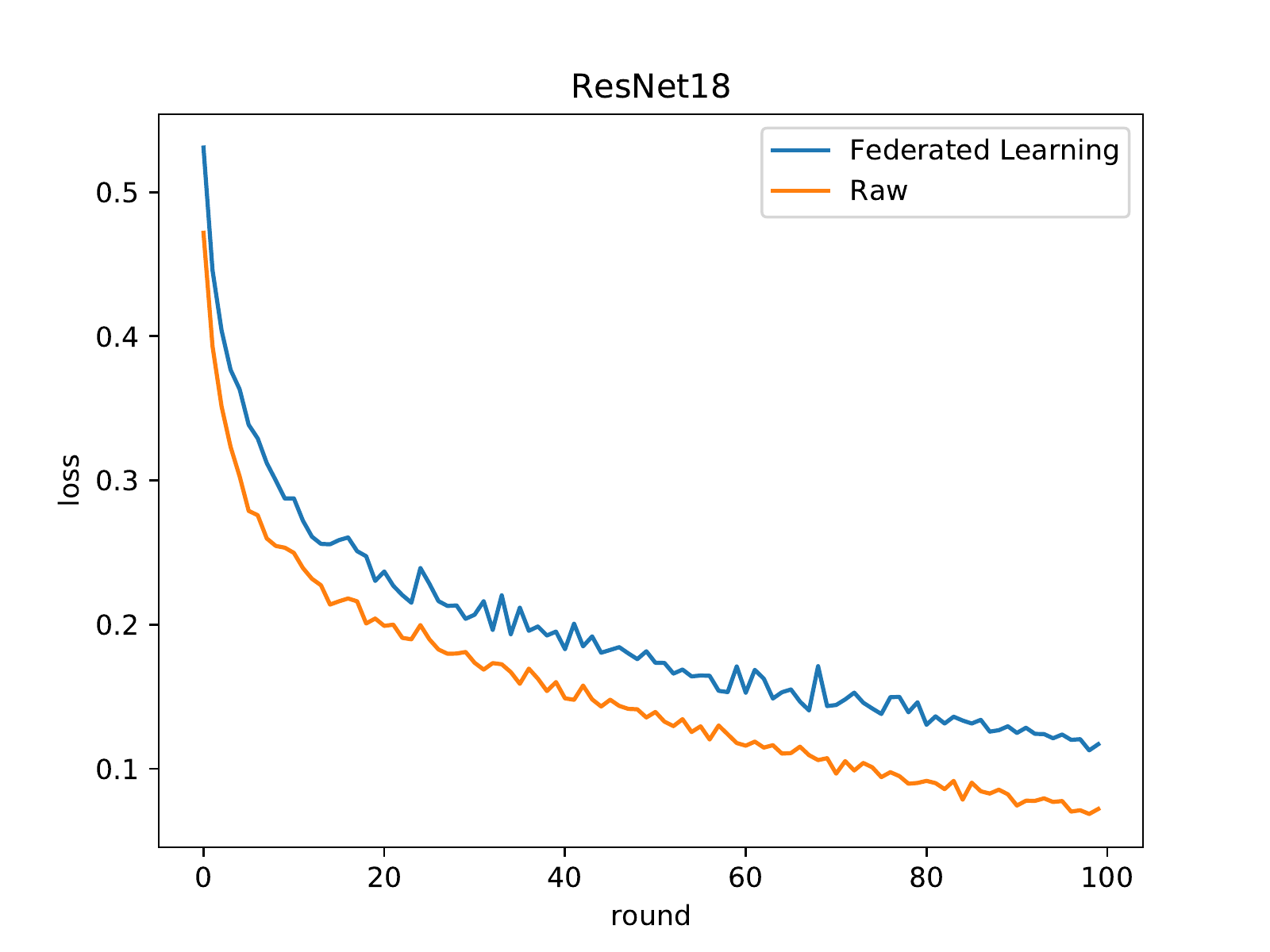}}
		\subfigure[ResNeXt FL]{\includegraphics[width=3.2in]{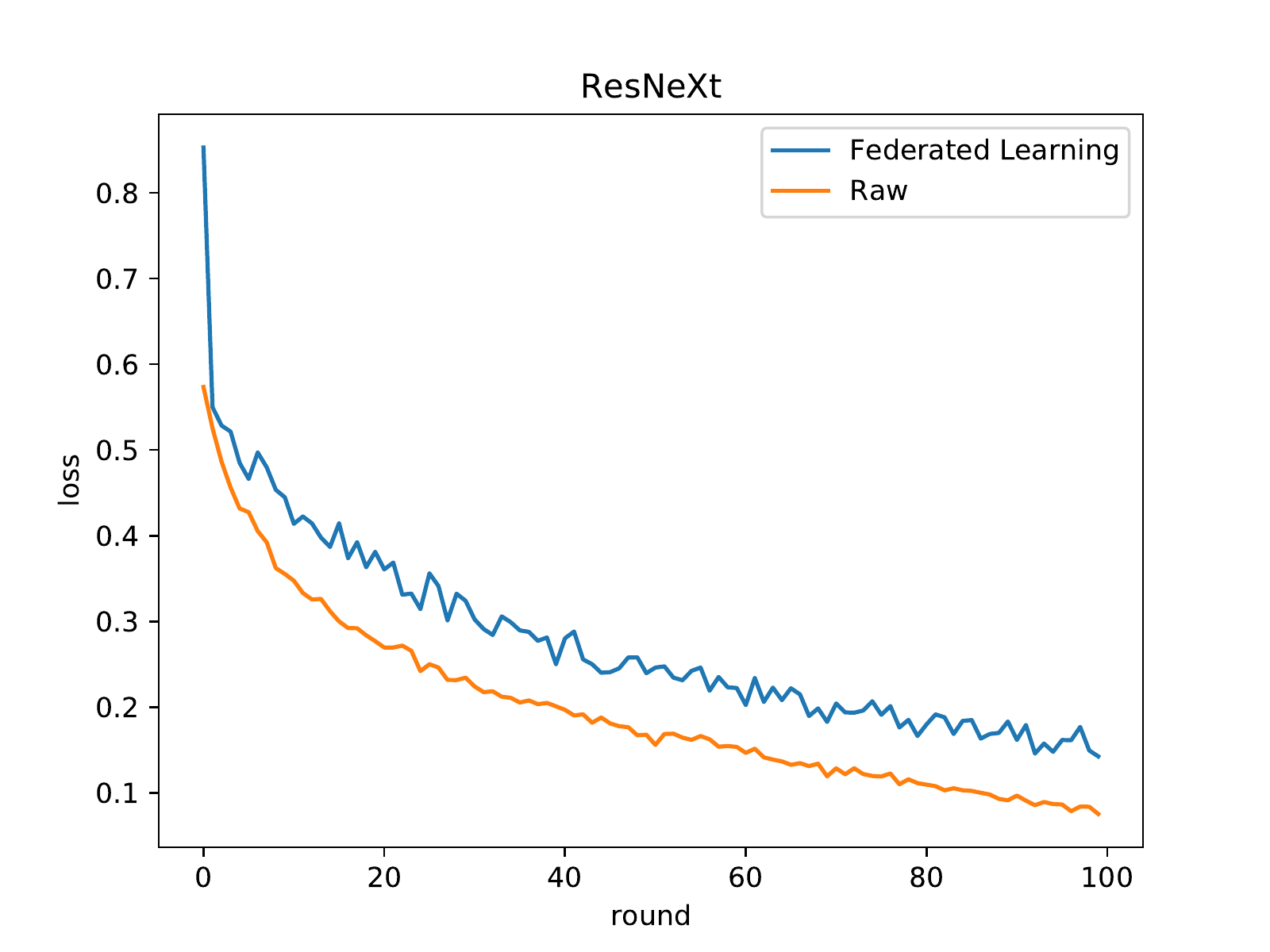}}
		\caption{Loss convergence speed comparison of whether to use the federated learning framework}
	\end{figure*}
	\subsection{Experimental Settings}
	\subsubsection{Model}
	The main task in the experiment is image classification. There are already many classic neural networks in this field, and there are models specifically designed for the recognition of COVID-19 pneumonia CXR images. Four models are used in the experiments.
	\begin{itemize}
		\item COVID-Net: A neural network specifically proposed to identify COVID-19 pneumonia CXR images utilizes PEPX to compress the network structure while preserving the network's performance to a large extent. At the same time, it has high sensitivity to the pneumonia characteristics of COVID-19.
		\item ResNet18: It is a residual neural network. An identity mapping layer is added to the ordinary neural network to make the network as deep as possible.To a certain extent, it can prevent the accuracy falling caused by overfitting due to the deepening of the network.
		\item ResNeXt: It is based on residual neural network using split-transform-merge strategy to convert single-core convolution into multi-core convolution, but the topology is the same as ResNet18.
		\item MobileNet-v2: It is a lightweight convolutional neural network. Unlike residual neural network, residual neural network uses a convolution kernel to first compress and extract features and then expand, but it expands and extracts more features and then compresses.
	\end{itemize}
	\subsubsection{Implementation and Training}
	Federated learning is a pseudo-distributed training completed on one machine in our experiment. For each agent, there is a separate model, which is reset to the updated central model after each central model update. The models are all implemented by PyTorch, and the training set and test set images are resized to (224,224) for model training. Each terminal agent uses the Adam optimizer with learning rate = 2e-5 and weight decay = 1e-7. The framework for federated learning is trained under the GPU acceleration of NVIDIA Tesla V100 (32GB) on Ubuntu 18.04 system. Other training-related parameters are shown in Table 1.
	\begin{table}[!htbp]
		\caption{Training-related parameters in the experiment}
		\centering 
		\begin{tabular}{ccc} 
			\toprule
			Parameter & \multicolumn{1}{p{3.69em}}{Value} & Description \\
			\midrule
			Agents number & 5     & \multicolumn{1}{p{15em}}{The number of agents} \\
			\multirow{3}{*}{Frac}  & \multirow{3}{*}{0.4}   & \multicolumn{1}{p{15em}}{The proportion of agents participating in the central model update for each round} \\
			Local epoch & 3     & \multicolumn{1}{p{15em}}{Epochs update per round} \\
			Local batch size & 10    & \multicolumn{1}{p{15em}}{Batch size update per round} \\
			Learning rate & 2.00E-05 & \multicolumn{1}{p{15em}}{Learning rate of optimizer} \\
			\multirow{2}{*}{Weight decay} & \multirow{2}{*}{1.00E-07} & \multicolumn{1}{p{15em}}{Decay of learning rate with training epoch} \\
			\bottomrule
		\end{tabular}
	\end{table}
	\subsection{Experimental Results and Analysis}
	During the experiment, the loss of several models can converge, and the decline of loss during the training process is shown in Fig 3.
	
	After using the same parameter training four models for 100 rounds, ResNet18 has the fastest convergence speed, and the highest accuracy rate (96.15\%, 91.26\%) on both the training set and the testing set. The ResNeXt convergence rate is closely followed, but the accuracy rate is not as good as the second-ranked COVID-Net. Although MobileNet-v2 has a loss value similar to COVID-Net, the accuracy rate on the testing set is not satisfactory.
	\begin{figure*}
		\centering
		\subfigure[CovidNet FL]{\includegraphics[width=1.6in]{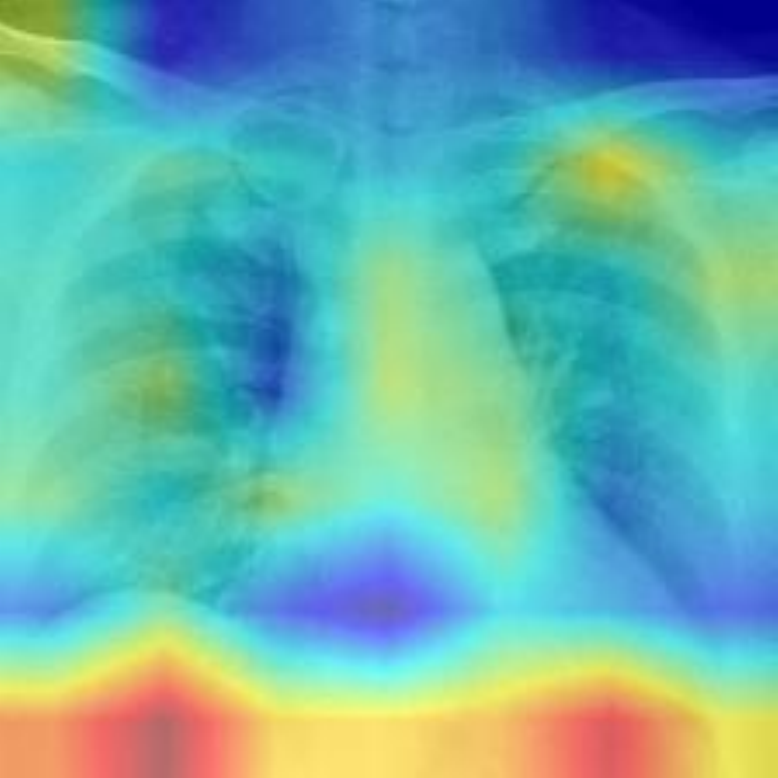}}
		\subfigure[MobileNet2 FL]{\includegraphics[width=1.6in]{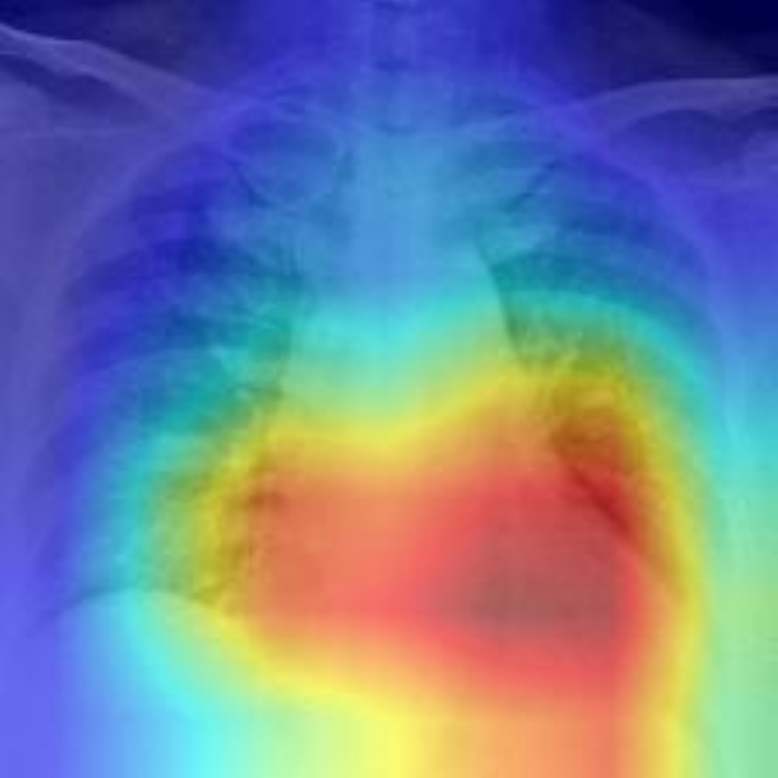}}
		\subfigure[ResNet18 FL]{\includegraphics[width=1.6in]{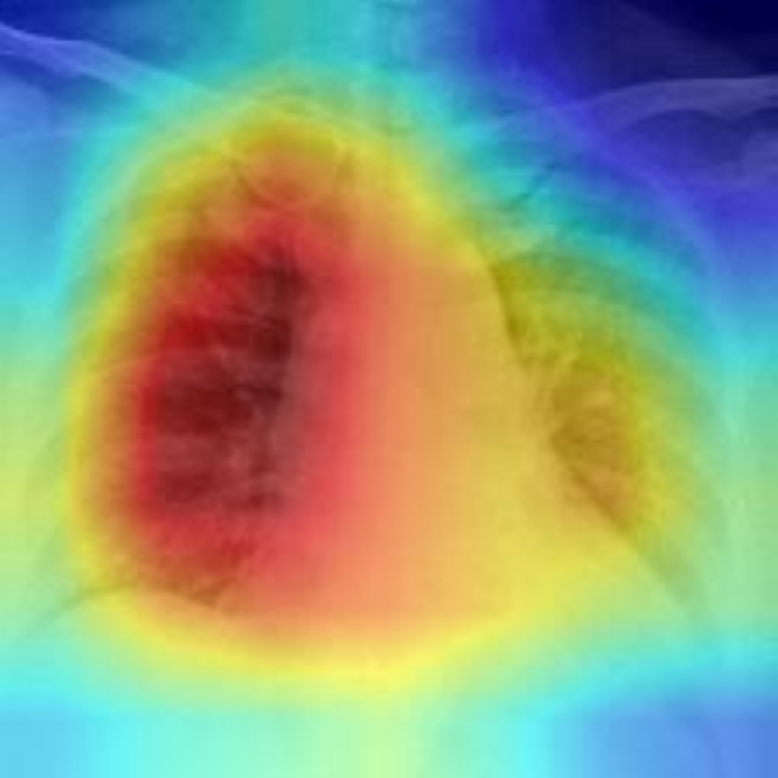}}
		\subfigure[ResNeXt FL]{\includegraphics[width=1.6in]{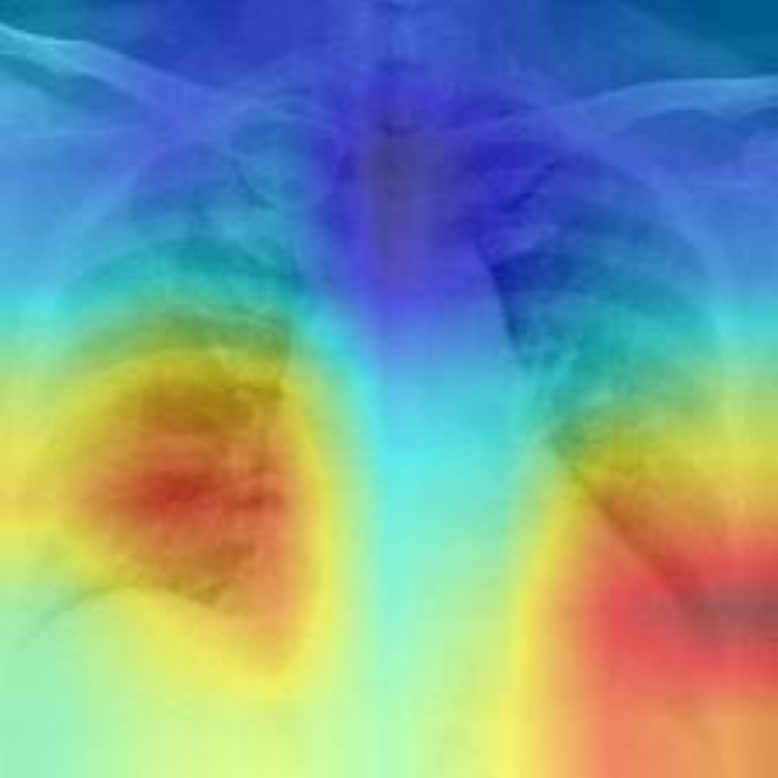}}
		\subfigure[CovidNet]{\includegraphics[width=1.6in]{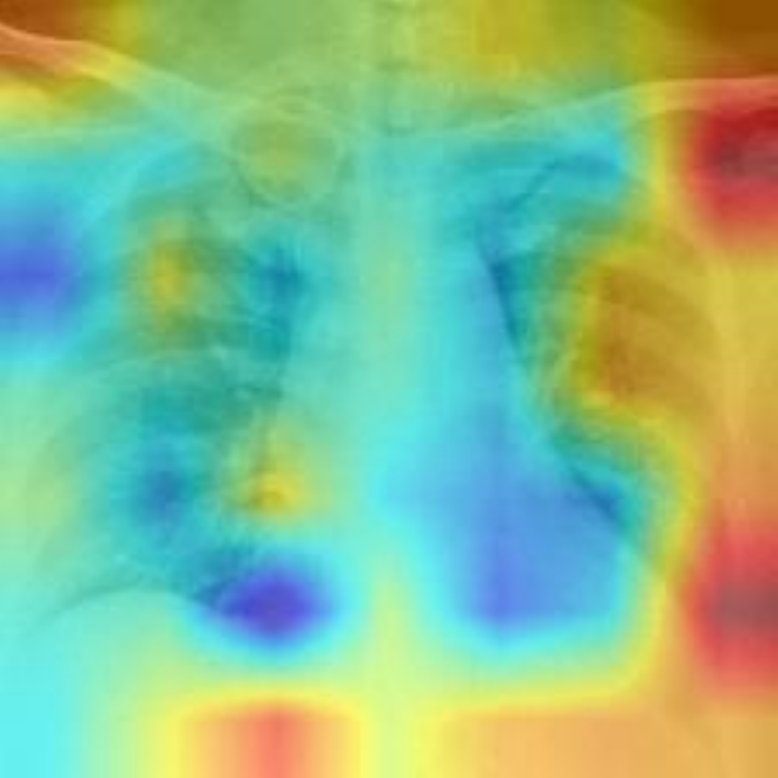}}
		\subfigure[MobileNet2]{\includegraphics[width=1.6in]{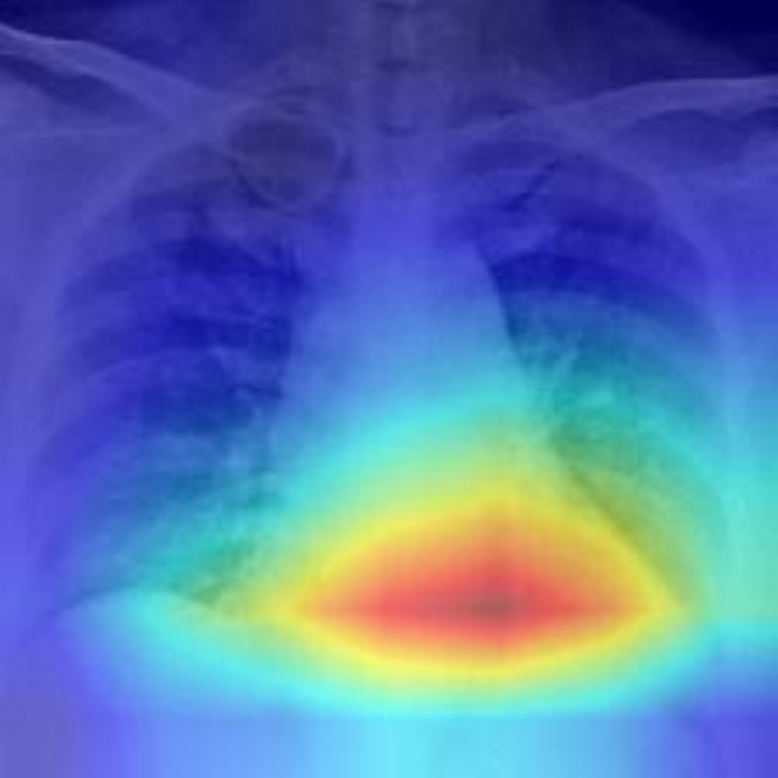}}
		\subfigure[ResNet18]{\includegraphics[width=1.6in]{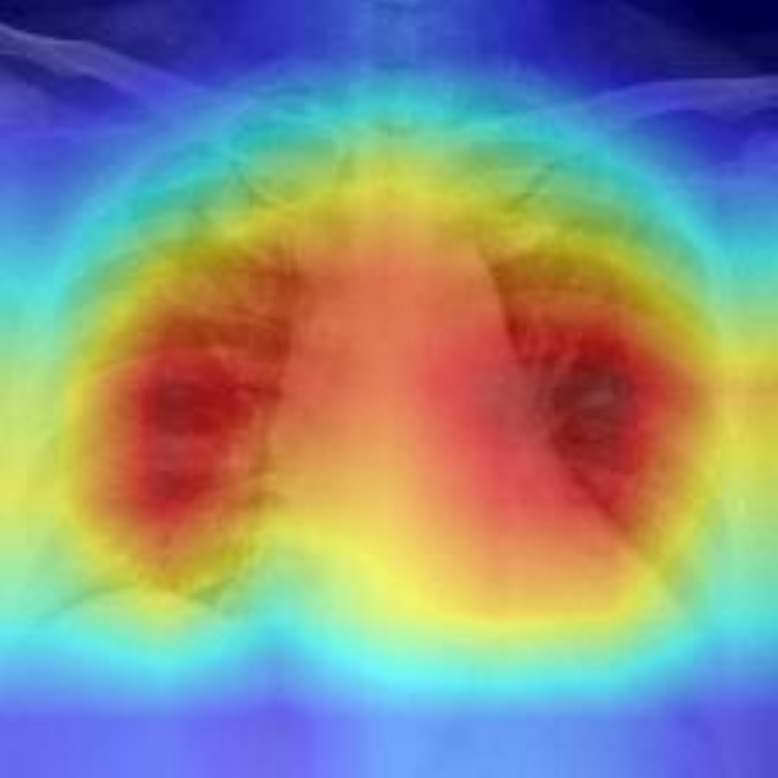}}
		\subfigure[ResNeXt]{\includegraphics[width=1.6in]{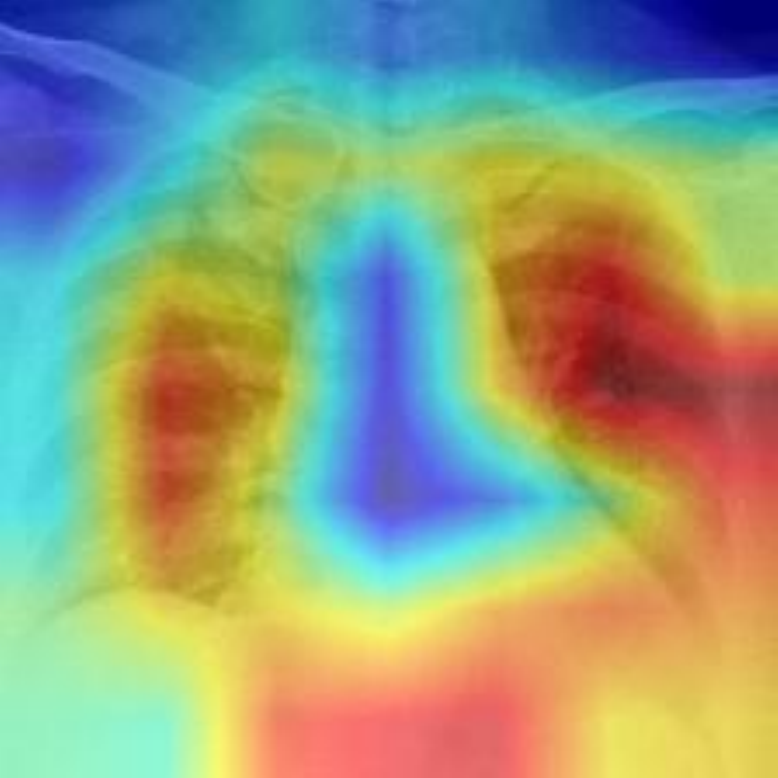}}
		\caption{Visual Explanations of the last convolution layer of each model, with COVID-19 label}
	\end{figure*}

	\begin{table}[!htbp]
		\centering
		\caption{Model sensitivity to data using federated learning framework}
		\begin{tabular}{ccc}
			\toprule
			Model & \multicolumn{1}{p{8em}<{\centering}}{Training Set} & \multicolumn{1}{p{8em}<{\centering}}{Testing Set} \\
			\midrule
			COVID-Net & \cellcolor[rgb]{ .796,  .796,  .796}92.40$\pm$0.004\% & \cellcolor[rgb]{ .631,  .631,  .631}89.17$\pm$0.015\% \\
			MobileNet\_v2 & \cellcolor[rgb]{ .745,  .745,  .745}91.16$\pm$0.005\% & \cellcolor[rgb]{ .502,  .502,  .502}86.83$\pm$0.017\% \\
			ResNet18 & \cellcolor[rgb]{ .949,  .949,  .949}\textbf{96.15$\pm$0.003\%} & \cellcolor[rgb]{ .749,  .749,  .749}\textbf{91.26$\pm$0.014\%} \\
			ResNeXt & \cellcolor[rgb]{ .886,  .886,  .886}94.66$\pm$0.004\% & \cellcolor[rgb]{ .698,  .698,  .698}90.37$\pm$0.015\% \\
			\bottomrule
		\end{tabular}%
		\label{tab:addlabel}%
	\end{table}%
	
	To explore the sensitivity of the models to each label, we counted the accuracy of each model for each label, and the performance results of the models are shown in Fig.4  and Tabel 3.
	\begin{table}[htbp]
		\centering
		\caption{Four models perplexity of each label}
		\begin{tabular}{p{6em}<{\centering}p{6em}<{\centering}p{6em}<{\centering}p{6em}<{\centering}}
			\toprule
			Model & Nomal & Non COVID-19 pneumonia & COVID-19 \\
			\midrule
			COVID-Net & \cellcolor[rgb]{ .902,  .902,  .902}96.47$\pm$0.004\% & \cellcolor[rgb]{ .733,  .733,  .733}88.24$\pm$0.009\% & \cellcolor[rgb]{ .506,  .506,  .506}51.04$\pm$0.05\% \\
			MobileNet-v2 & \cellcolor[rgb]{ .863,  .863,  .863}94.87$\pm$0.005\% & \cellcolor[rgb]{ .725,  .725,  .725}87.20$\pm$0.009\% & \cellcolor[rgb]{ .502,  .502,  .502}50.26$\pm$0.05\% \\
			ResNet18 & \cellcolor[rgb]{ .949,  .949,  .949}\textbf{98.16$\pm$0.003\%} & \cellcolor[rgb]{ .835,  .835,  .835}\textbf{93.91$\pm$0.006\%} & \cellcolor[rgb]{ .6,  .6,  .6}66.32$\pm$0.047\% \\
			ResNeXt & \cellcolor[rgb]{ .894,  .894,  .894}96.18$\pm$0.004\% & \cellcolor[rgb]{ .804,  .804,  .804}92.66$\pm$0.007\% & \cellcolor[rgb]{ .643,  .643,  .643}\textbf{73.58$\pm$0.044\%} \\
			\bottomrule
		\end{tabular}%
		\label{tab:addlabel}%
	\end{table}%
	
	At the same time, we compared the training results without the federated learning framework with those using the federated learning framework. Because round=100 and local\_epoch=3 in the federated learning parameters, we set epoch=300 when training the model separately, so as to compare with the loss convergence during the federated learning training process, as shown in Fig.5. It was found that the loss convergence rate caused by the use of federated learning decreased slightly. The result of training accuracy of a single network is shown in Table 4.
	
	\begin{table}[!htbp]
		\centering
		\caption{Loss convergence speed comparison of whether to use the federal learning framework}
		\begin{tabular}{p{7em}<{\centering}cc}
			\toprule
			Model & \multicolumn{1}{p{8em}<{\centering}}{Training Set} & \multicolumn{1}{p{8em}<{\centering}}{Testing Set} \\
			\midrule
			COVID-Net & \cellcolor[rgb]{ .816,  .816,  .816}94.50$\pm$0.004\% & \cellcolor[rgb]{ .573,  .573,  .573}90.06$\pm$0.015\% \\
			MobileNet-v2 & \cellcolor[rgb]{ .8,  .8,  .8}94.10$\pm$0.004\% & \cellcolor[rgb]{ .502,  .502,  .502}88.98$\pm$0.015\% \\
			ResNet18 & \cellcolor[rgb]{ .949,  .949,  .949}\textbf{98.06$\pm$0.002\%} & \cellcolor[rgb]{ .639,  .639,  .639}91.07$\pm$0.014\% \\
			ResNeXt & \cellcolor[rgb]{ .933,  .933,  .933}97.66$\pm$0.003\% & \cellcolor[rgb]{ .651,  .651,  .651}\textbf{91.26$\pm$0.014\%} \\
			\bottomrule
		\end{tabular}%
		\label{tab:addlabel}%
	\end{table}%
	 As presented in Fig.5, we compared the training procedure based on federated learning with the training procedure without federated learning. MobileNet-V2 has a larger accuracy gap between the FL based approach and the individual approach while ResNet18 has a smaller accuracy gap. If considering the number of parameters at the same time, MobileNet and ResNet18 have a higher performance. MoblieNet has the fewest parameters and the lowest accuracy. ResNet18 has the second fewest parameters and the highest accuracy. 
	 
	We used the Grad-CAM++ method to perform Visual Explanations on the models, and the results are shown in Fig 6-8. From left to right in each figure are the COVID-Net, ResNet18, ResNeXt, and MobileNet-v2 models. The first line is the result of training using the federated learning framework, and the second line is the result of training without the federated learning framework.
		\begin{figure*}
		\centering
		\subfigure[CovidNet FL]{\includegraphics[width=1.6in]{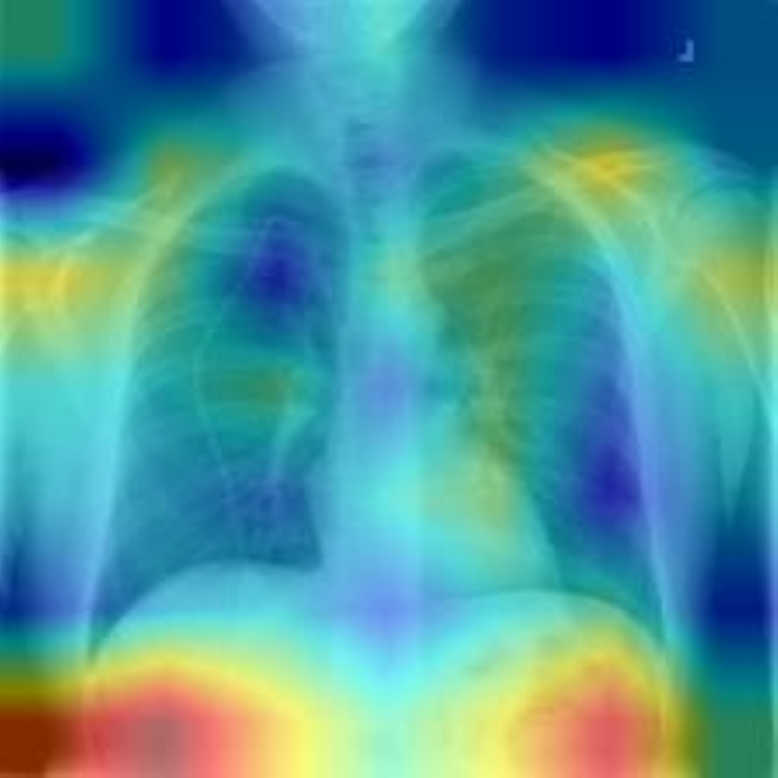}}
		\subfigure[MobileNet2 FL]{\includegraphics[width=1.6in]{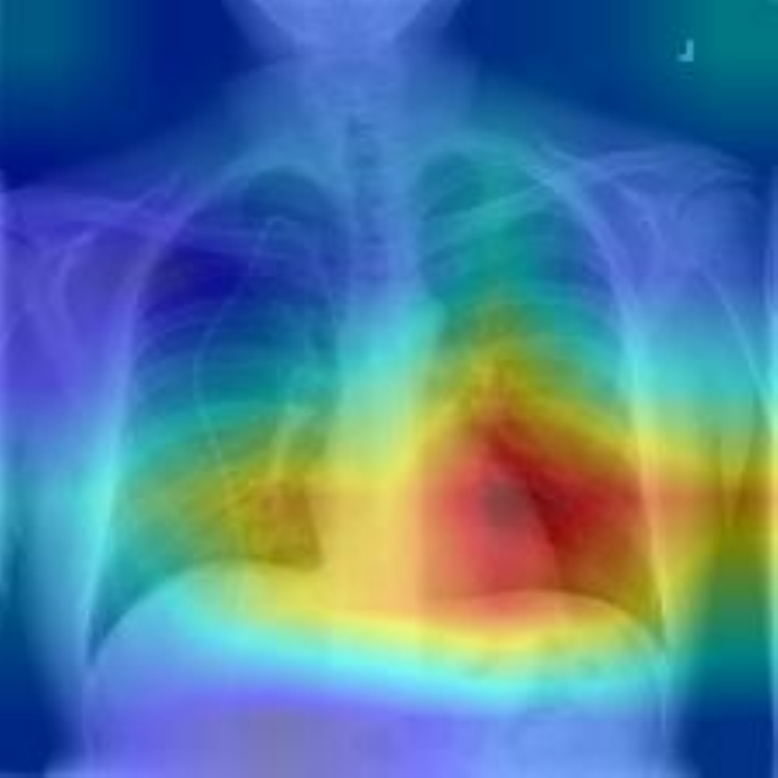}}
		\subfigure[ResNet18 FL]{\includegraphics[width=1.6in]{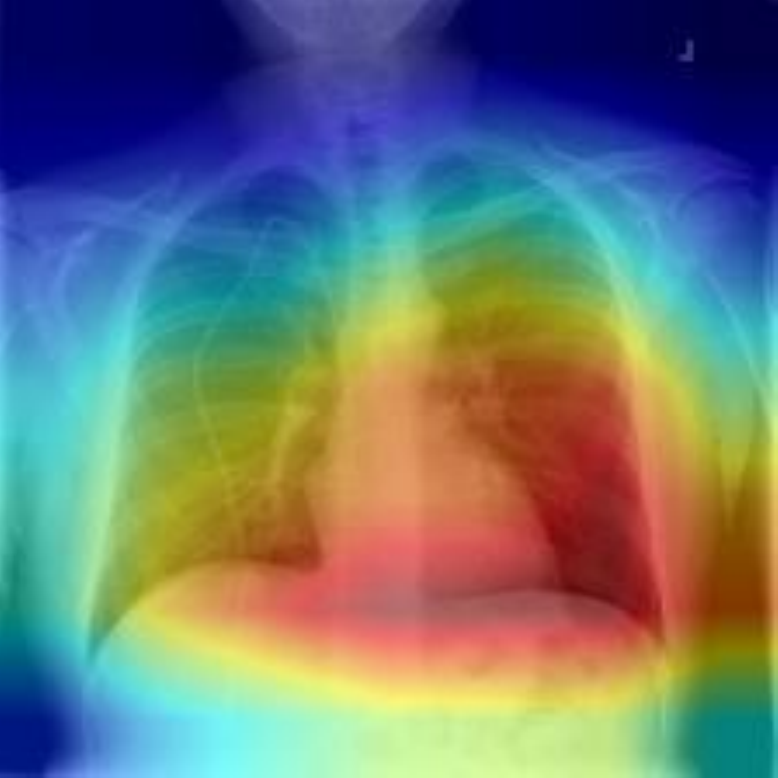}}
		\subfigure[ResNeXt FL]{\includegraphics[width=1.6in]{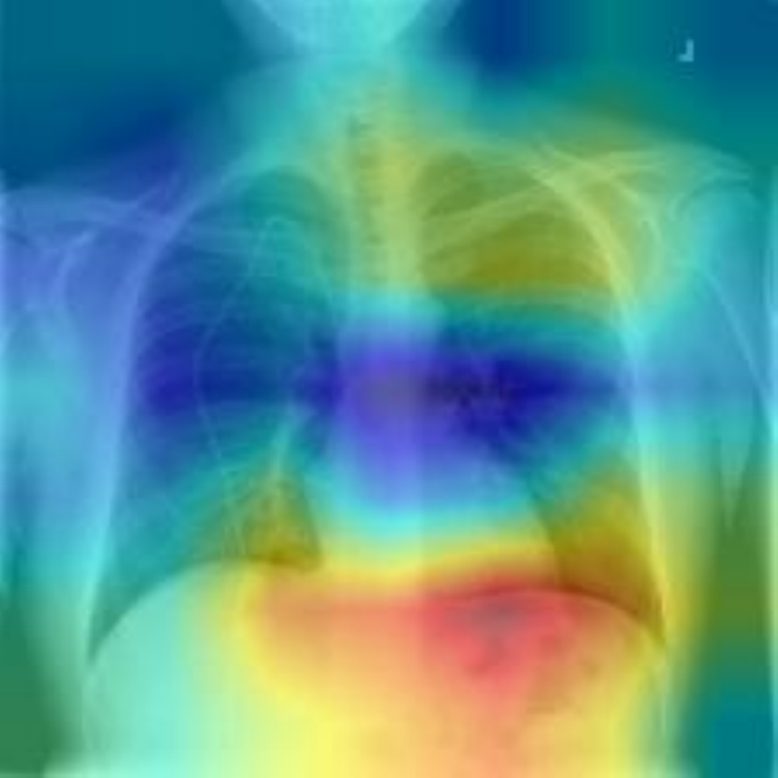}}
		\subfigure[CovidNet]{\includegraphics[width=1.6in]{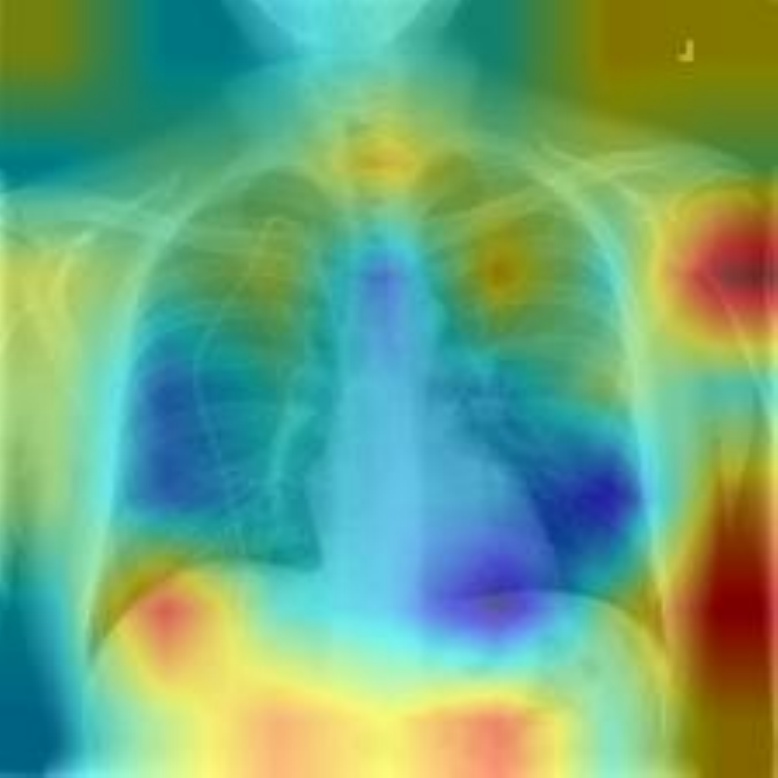}}
		\subfigure[MobileNet2]{\includegraphics[width=1.6in]{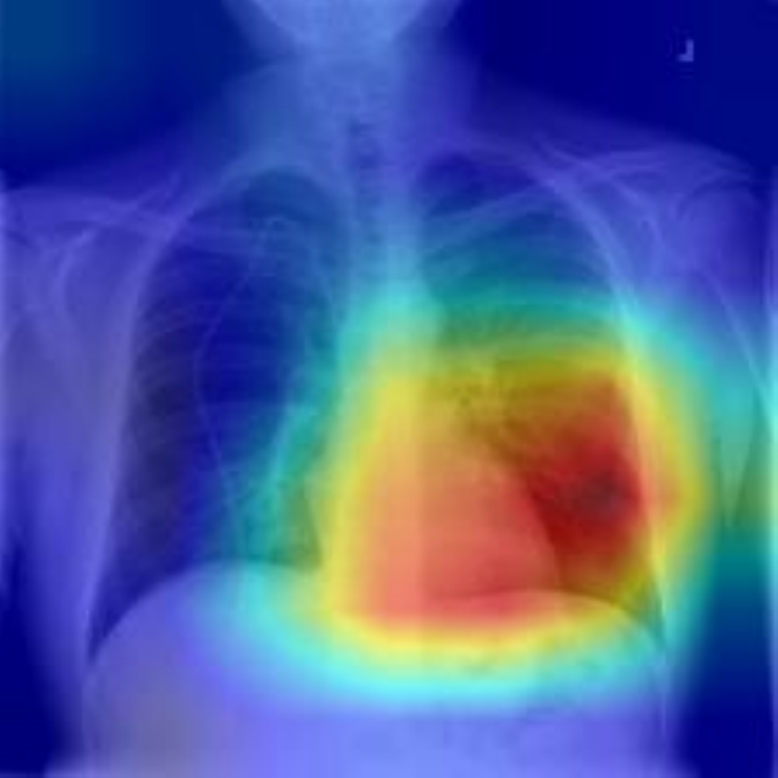}}
		\subfigure[Resnet18]{\includegraphics[width=1.6in]{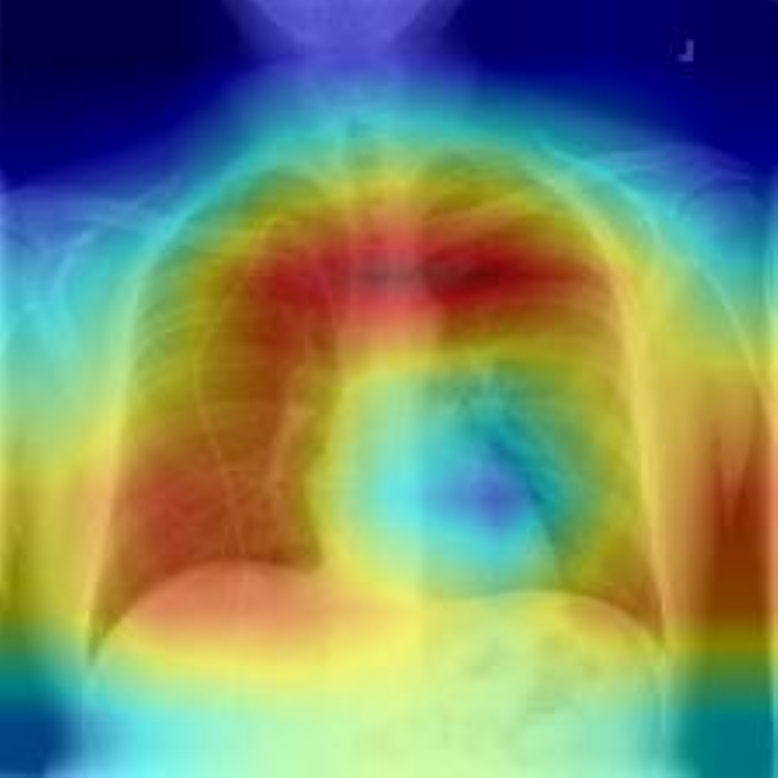}}
		\subfigure[ResNeXt]{\includegraphics[width=1.6in]{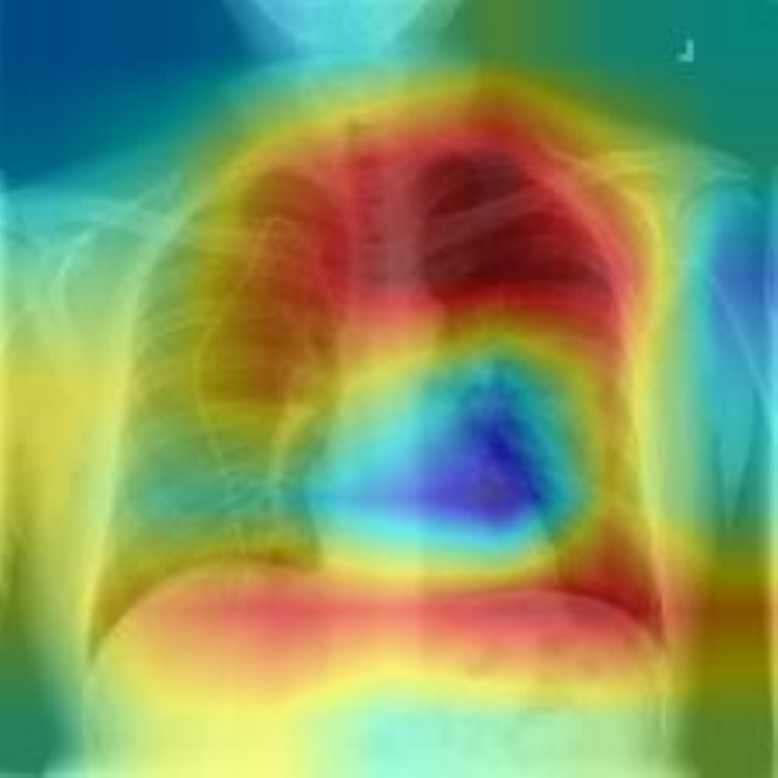}}
		\caption{Visual Explanations of the last convolution layer of each model, with normal label}
	\end{figure*}
	\begin{figure*}
		\centering
		\subfigure[CovidNet FL]{\includegraphics[width=1.6in]{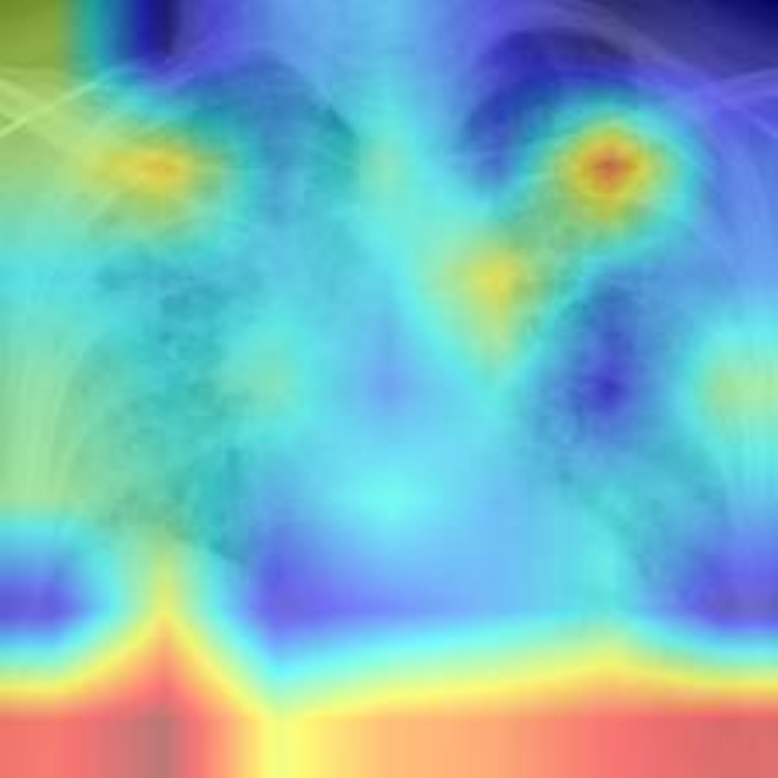}}
		\subfigure[Mobile\textsc{nN}et2 FL]{\includegraphics[width=1.6in]{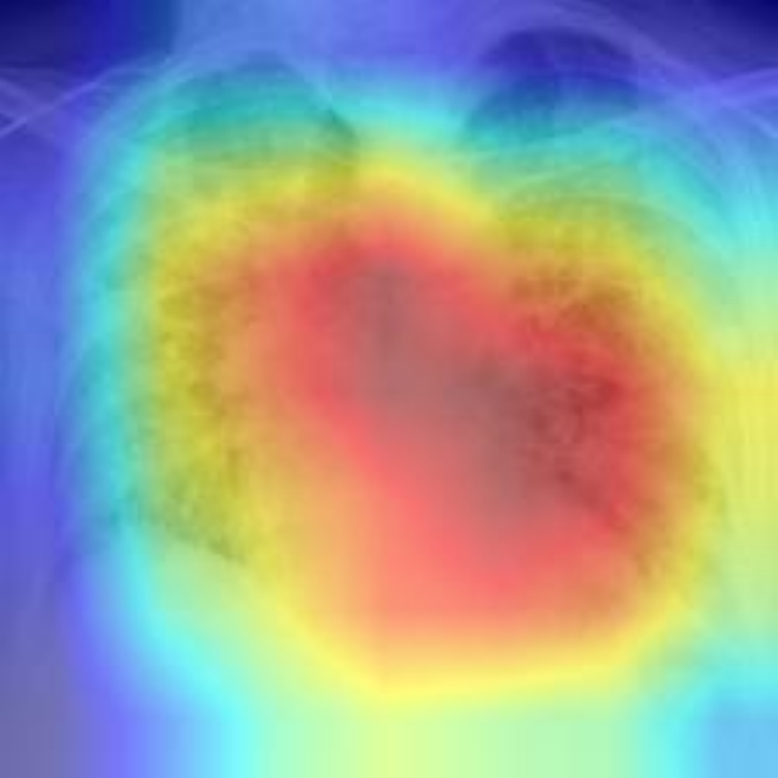}}
		\subfigure[ResNet18 FL]{\includegraphics[width=1.6in]{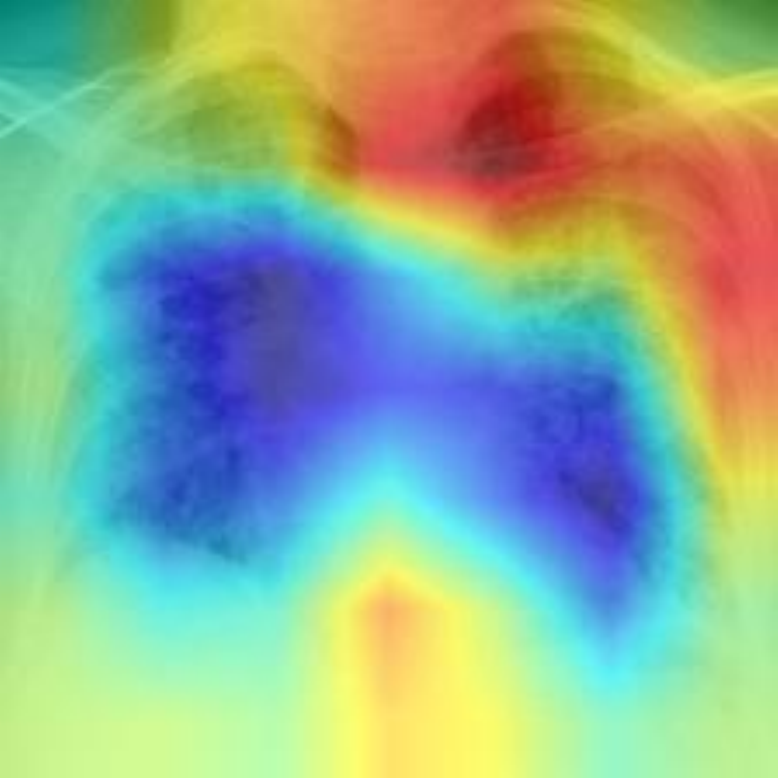}}
		\subfigure[ResNeXt FL]{\includegraphics[width=1.6in]{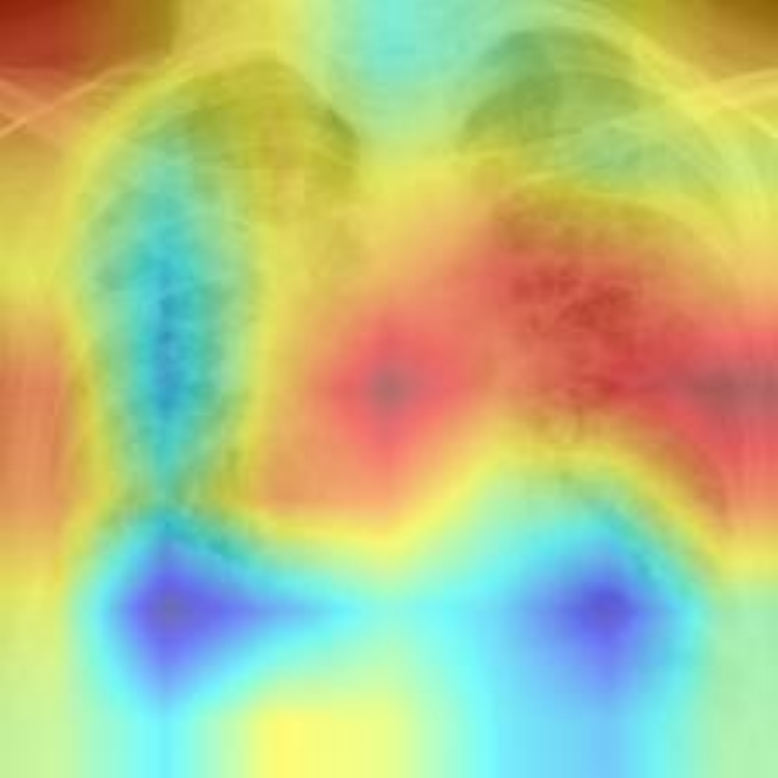}}
		\subfigure[CovidNet]{\includegraphics[width=1.6in]{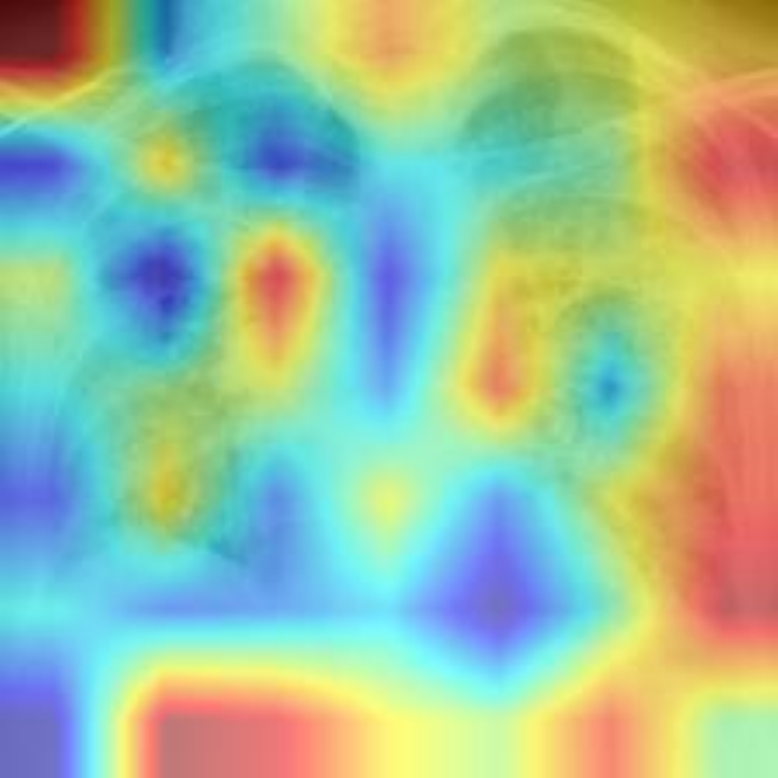}}
		\subfigure[MobileNet2]{\includegraphics[width=1.6in]{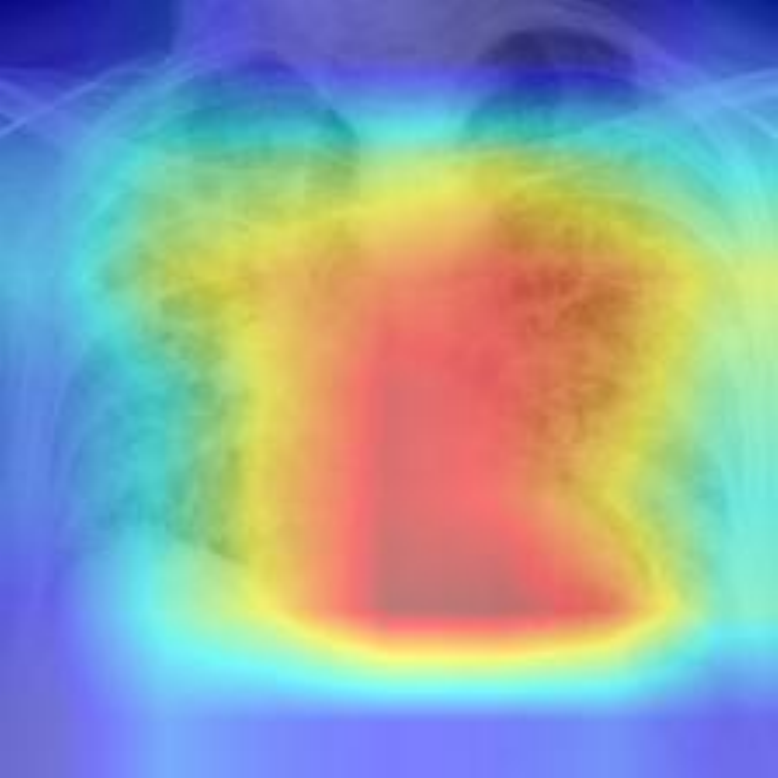}}
		\subfigure[ResNet18]{\includegraphics[width=1.6in]{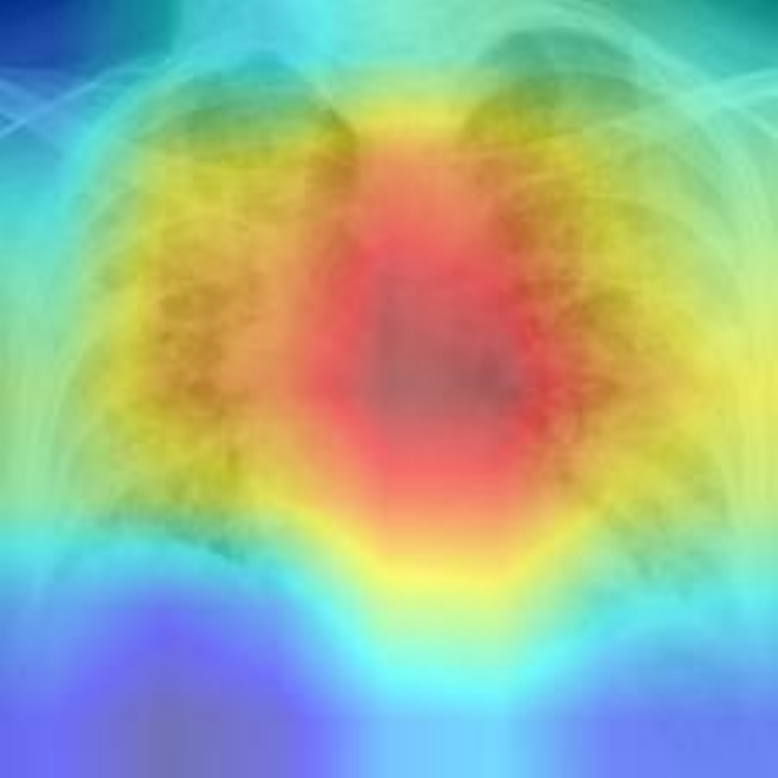}}
		\subfigure[ResNeXt]{\includegraphics[width=1.6in]{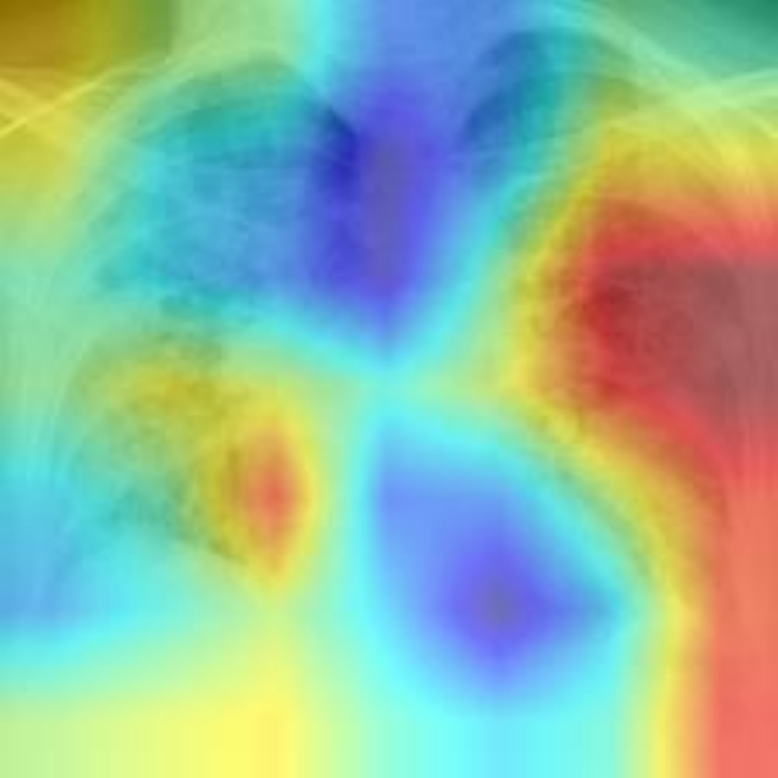}}
		\caption{Visual Explanations of the last convolution layer of each model, with pneumonia label}
	\end{figure*}
    \section{Conclusion}
    In the work, we conducted experiments on COVID-19 identification with CXR images based on the federated learning framework. We conducted CXR images trainning with four different models: MobileNet, ResNet18, MoblieNet and COVID-Net, and comparison experiment between training with federated learning framework and training without federated learning framework.The experimental results show that ResNet18 has the best performance both in training with FL and without FL. ResNeXt has the best performance in images with COVID-19 labels. MoblieNet has the fewest number of parameters. Therefore, the work indicates that ResNeXt and ResNet18 are better chosen for COVID-19 identification among the four popular models.
	
	\bibliographystyle{IEEEtran}
	\bibliography{citations}
\end{document}